\documentclass[aps,prx,twocolumn,superscriptaddress,10pt]{revtex4-2}
\usepackage[colorlinks, allcolors=Blue, linktocpage]{hyperref}
\usepackage[dvipsnames]{xcolor}
\usepackage{amsmath,amssymb,graphicx,CJK}
\usepackage{newtxtext,newtxmath,mathrsfs,multirow,enumitem}
\usepackage[boxsize=.5em, aligntableaux=center]{ytableau}
\usepackage[T1]{fontenc}
\usepackage[utf8]{inputenc}

\setlist{align=parleft,leftmargin=0pt,itemindent=2.5em,labelsep=0em,labelwidth=1.5em,nosep}

\begin{document}

\begin{CJK*}{UTF8}{bsmi}

\title{\texorpdfstring{\boldmath}{} The \texorpdfstring{$\mathrm{SO}(5)$}{SO(5)} Deconfined Phase Transition under the Fuzzy Sphere Microscope:\texorpdfstring{\\}{} Approximate Conformal Symmetry, Pseudo-Criticality, and Operator Spectrum}

\author{Zheng Zhou (周正)}
\affiliation{Perimeter Institute for Theoretical Physics, Waterloo, Ontario N2L 2Y5, Canada}
\affiliation{Department of Physics and Astronomy, University of Waterloo, Waterloo, Ontario N2L 3G1, Canada}

\author{Liangdong Hu}
\affiliation{Westlake Institute of Advanced Study, Westlake University, Hangzhou 310024, China}

\author{W. Zhu}
\affiliation{Westlake Institute of Advanced Study, Westlake University, Hangzhou 310024, China}

\author{Yin-Chen He}
\affiliation{Perimeter Institute for Theoretical Physics, Waterloo, Ontario N2L 2Y5, Canada}

\begin{abstract}
    The deconfined quantum critical point (DQCP) is an example of phase transitions beyond the Landau symmetry breaking paradigm that attracts wide interest. However, its nature has not been settled after decades of study. In this paper, we apply the recently proposed fuzzy sphere regularization to study the $\mathrm{SO}(5)$ non-linear sigma model (NL$\sigma$M) with a topological Wess-Zumino-Witten term, which serves as a dual description of the DQCP with an exact $\mathrm{SO}(5)$ symmetry. We demonstrate that the fuzzy sphere functions as a powerful microscope, magnifying and revealing a wealth of crucial information about the DQCP, ultimately paving the way towards its final answer. In particular, through exact diagonalization, we provide clear evidence that the DQCP exhibits approximate conformal symmetry. The evidence includes the existence of a conserved $\mathrm{SO}(5)$ symmetry current, a stress tensor, and integer-spaced levels between conformal primaries and their descendants. Most remarkably, we have identified 19 conformal primaries and their 82 descendants.  Furthermore, by examining the renormalization group flow of the lowest symmetry singlet, we demonstrate that the DQCP is more likely pseudo-critical, with the approximate conformal symmetry plausibly emerging from nearby complex fixed points. Our computed primary spectrum also has important implications, including the conclusion that the $\mathrm{SO}(5)$ DQCP cannot describe a direct transition from the N\'eel to valence bond solid phase on the honeycomb lattice.
\end{abstract}

\maketitle

\end{CJK*}

\section{Introduction}

The study of universal properties of quantum phase transitions has been a central task within the condensed matter physics community \cite{Sachdev2011Quantum}. While Landau's spontaneous symmetry-breaking paradigm has provided insights into many criticalities, researchers have also identified exotic quantum phase transitions that defy this paradigm. One famous example is the deconfined quantum critical point (DQCP)~\cite{Senthil2004Deconfined, Senthil2004Quantum}, initially proposed to describe a direct quantum phase transition between the N\'eel anti-ferromagnet (AFM) and the valence bond solid (VBS) on a square lattice~\cite{Read1989SpinPeierlsB,Read1989SpinPeierlsL,Murthy1990Instantons}. DQCP, besides being one of the pioneering phase transitions beyond Landau symmetry breaking, has led to numerous theoretical surprises, including the emergence of $\mathrm{SO}(5)$ symmetry \cite{Nahum2015Emergent} and the conjectural duality between different strongly interacting 3D (i.e., $(2+1)$D) gauge theories \cite{Wang2017Deconfined}.

Since its proposal, DQCP has undergone extensive studies in various models over the past two decades \cite{Nahum2015Emergent,Motrunich2004Emergent,Sandvik2007Evidence,Melko2008Scaling,Jiang2008DQCPFirstorder,Kuklov2008Deconfined,Sandvik2010Continuous,Lou2009Antiferromagnetic,Sandvik2010Continuous,Banerjee2010Impurity,Kaul2012Lattice,Kaul2012Quantum,Bartosch2013Corrections,Chen2013Deconfined,Sumiran2013Neel,Nahum2015Deconfined,Sreejith2015Monopoles,Sreejith2014Monopoles,Shao2016Quantum,Sato2017Dirac,Serna2019Emergence,Liu2019Superconductivity,Zhao2020Multicritical,Wang2021Phases} (also see a recent review~\cite{Senthil2023DQCPReview}). However, its nature remains controversial. Monte Carlo simulations of DQCP in different models have shown no signals of discontinuity, thus suggesting a continuous phase transition. However, abnormal scaling behaviors have been observed \cite{Nahum2015Emergent,Nahum2015Deconfined}. Moreover, the critical exponents obtained by Monte Carlo simulations violate the rigorous bounds from the conformal bootstrap method \cite{Poland2019Conformal,Nakayama2016Necessary}. Several possibilities have been proposed to reconcile these tensions. One possibility is that DQCP represents a continuous quantum phase transition that does not exhibit emergent conformal symmetry. Alternatively, it is possible that previous Monte Carlo analyses failed to obtain precise critical exponents due to the presence of abnormal scaling behavior~\cite{Sandvik2010Continuous,Bartosch2013Corrections,Shao2016Quantum,Nahum2015Deconfined}. Among these possibilities, a particularly intriguing proposal is the concept of pseudo-criticality~\cite{Wang2017Deconfined}, also known as walking behavior caused by a complex fixed point~\cite{Gorbenko2018Walking,Gorbenko2018Walking2}. According to this proposal, DQCP manifests as a weakly first-order transition with a tiny gap and a large correlation length. In contrast to conventional first-order phase transitions, a pseudo-critical system demonstrates behavior closely resembling a continuous transition, adhering to universal behaviors, especially at energy scales (e.g., finite temperature in real experiments) larger than its small energy gap.

Theoretically, a pseudo-critical system resides near complex fixed points within the unphysical (non-unitary) parameter space. Under the renormalization group (RG) flow, the system slowly walks in the shadow of complex fixed points, resulting in approximate scaling and conformal symmetry~\cite{Wang2017Deconfined,Gorbenko2018Walking,Gorbenko2018Walking2}. A notable example of pseudo-criticality is observed in a famous weakly first-order transition, i.e., the order-disorder transition of the 2D ($(1+1)$D) 5-state Potts model~\cite{Nienhuis1979Potts}. Numerical investigations have indeed revealed approximate conformal symmetry in this model~\cite{Ma2019Potts}. However, despite the plausibility of pseudo-criticality as a scenario consistent with most numerical findings regarding DQCP~\cite{Ma2020Theory,Nahum2020WZW}, direct evidence remains elusive.

In general, understanding 3D interacting phase transitions such as DQCP requires non-perturbative tools like numerical methods. Historically, Monte Carlo simulation of lattice models has been the primary reliable numerical method for studying 3D phase transitions. However, this approach faces several challenges when tackling problems like DQCP. Firstly, Monte Carlo simulations do not provide direct information regarding the emergent conformal symmetry. Secondly, they heavily rely on finite-size extrapolation of physical observables such as correlators to extract critical exponents, limiting access to only a small number of critical exponents. Moreover, subtle issues may arise if there is abnormal scaling behavior in the pseudo-critical system. Lastly, the similarity in behavior between pseudo-critical and true critical systems makes it exceedingly difficult to distinguish them, despite a few proposals~\cite{Emidio2021Diagnosing,Zhao2022Scaling} that currently lack a well-founded theoretical foundation.

A novel approach, called fuzzy sphere regularization, has recently emerged as a powerful method for studying critical phenomena in 3D (i.e., $(2+1)$D)~\cite{Zhu2023Uncovering}. It involves investigating the $(2+1)$D quantum phase transition on the geometry of $S^2\times \mathbb{R}$ using the fuzzy (non-commutative) sphere~\cite{Madore1992Fuzzy}. This approach offers distinct advantages over traditional lattice model-based methods, including the direct observation of emergent conformal symmetry and the efficient extraction of critical data, such as critical exponents, by employing the state-operator correspondence, without relying on finite-size extrapolation. A key feature of the fuzzy sphere scheme is the state-operator correspondence~\cite{Cardy1984Conformal, Cardy1985Universal}, which allows easy access to information such as the scaling dimensions of many operators. Specifically, there is a one-to-one correspondence between the eigenstates $|k\rangle$ of the CFT quantum Hamiltonians on the sphere and the CFT operators. Moreover, the energy gaps $\delta E_k$ are proportional to the scaling dimensions $\Delta_k$ of the CFT operators
\begin{equation}
    \delta E_k=E_k-E_0=\mathrm{constant}\times\Delta_k,
    \label{eq:2c_state_operator}
\end{equation}
where the scale factor is model and size-dependent. The power of this approach has been demonstrated in the context of the 3D Ising transition, where the presence of emergent conformal symmetry has been convincingly established~\cite{Zhu2023Uncovering}. Moreover, accurate and efficient determinations of 15 primary operators (i.e., independent critical exponents)~\cite{Zhu2023Uncovering}, 13 operator product expansion (OPE) coefficients~\cite{Hu2023Operator}, and several four-point correlators~\cite{Han2023Conformal} have been achieved as well. Therefore, the fuzzy sphere can serve as a powerful microscope for studying 3D critical phenomena, magnifying and revealing crucial information that is inaccessible through other approaches.

In this paper, using the fuzzy sphere microscope, we provide direct evidence that the $\mathrm{SO}(5)$ DQCP is pseudo-critical with an approximate conformal symmetry. Specifically, we investigate the 3D $\mathrm{SO}(5)$ non-linear sigma model (NL$\sigma$M) with a level-1 Wess-Zumino-Witten (WZW) term, which serves as one of the dual descriptions of the $\mathrm{SO}(5)$ DQCP~\cite{Nahum2015Emergent, Wang2017Deconfined}. A similar model has been previously studied on the torus using determinant Monte Carlo methods~\cite{Ippoliti2018Half, Wang2021Phases}. Throughout a wide range of interaction strengths, we observe an approximate conformal symmetry in the excitation spectrum of the Hamiltonian, confirmed by the identification of the conserved $\mathrm{SO}(5)$ symmetry current, stress tensor, and the (approximately) integer-spaced levels between various conformal primaries and their descendants. Interestingly, as we vary the system size, we find that the renormalization group (RG) flow supports the scenario of pseudo-criticality. In particular, we observe the lowest symmetry singlet flowing from being slightly irrelevant to slightly relevant, which is a characteristic feature of pseudo-criticality. The scaling dimensions of other primaries also exhibit size and parameter dependences that are quantitatively consistent with the prediction of pseudo-criticality. Furthermore, we identify and calculate the scaling dimensions of various primary operators in the operator spectrum, some of which are crucial for understanding the physics of the DQCP. For instance, our estimated critical exponent $\eta$ for the $\mathrm{SO}(5)$ order parameter is consistent with previous Monte Carlo estimations. Additionally, we find that the lowest parity-odd $\mathrm{SO}(5)$ singlet is highly irrelevant with a scaling dimension of approximately $\Delta\approx 5.4$. If this operator were relevant, it would drive the DQCP towards a chiral spin liquid, potentially playing a role in interesting phenomena observed in real materials~\cite{samajdar2019enhanced}. We also identify a relevant $6\pi$-monopole (in the language of the CP$^1$ model~\cite{Murthy1990Instantons}), indicating that the N\'eel-VBS transition on the honeycomb lattice cannot be described by the $\mathrm{SO}(5)$ DQCP~\cite{Sumiran2013Neel}. Conversely, the $8\pi$-monopole is found to be irrelevant, supporting the original conjecture of a stable DQCP on the square lattice.

The paper is organized as follows. In Sec.~\ref{sec:2a}, we explain the Hamiltonian of the non-linear sigma model on the lowest Landau level and fuzzy sphere. We then discuss the possible scenarios of the RG flow, in particular, the conformal window and pseudo-criticality in Sec.~\ref{sec:2b}; we discuss the quantitative predictions for pseudo-criticality from conformal perturbation in Sec.~\ref{sec:2b_1}; following that, in Sec.~\ref{sec:2c}, we discuss the relation with the original N\'eel-VBS DQCP. We then present our numerical results in Sec.~\ref{sec:3}. In particular, we provide strong evidence for approximate conformal symmetry and support the scenario of pseudo-criticality in Sec.~\ref{sec:3c}; in Sec.~\ref{sec:3d}, we discuss the operator spectrum and its physical consequence; in Sec.~\ref{sec:3d_1}, we analyze the drift of scaling dimensions and show that the results are consistent with the prediction of pseudo-criticality; in Sec.~\ref{sec:3e}, we calculate the correlation functions and OPE coefficients. Finally, we present a summary and discussion in Sec.~\ref{sec:4}. The appendices contain more discussion about the formalism and detailed numerical data of the spectra.

\section{Deconfined phase transition on the fuzzy sphere}

\subsection{Model}
\label{sec:2a}

The deconfined quantum critical point (DQCP) has multiple dual theoretical descriptions~\cite{Wang2017Deconfined}, one of which involves the 3D $\mathrm{SO}(5)$ non-linear sigma model (NL$\sigma$M) with a level-1 Wess-Zumino-Witten (WZW) term~\cite{Nahum2015Emergent}. It has been found that this $\mathrm{SO}(5)$ NL$\sigma$M can be naturally realized with electrons in the half-filled lowest Landau level (LLL), where an intriguing aspect is that the UV Hamiltonian has an exact $\mathrm{SO}(5)$ symmetry~\cite{Ippoliti2018Half,Wang2021Phases}. Our fuzzy sphere model is a spherical realization of this proposal, albeit formulated in a slightly different form.

The target space of the $\mathrm{SO}(5)$ NL$\sigma$M is $S^4 \cong \frac{\mathrm{SO}(5)}{\mathrm{SO}(4)} \cong \frac{\mathrm{Sp}(2)}{\mathrm{Sp}(1)\times \mathrm{Sp}(1)}$. Here, we adopt the convention that $\mathrm{Sp}(1)\cong \mathrm{SU}(2)$ and $\mathrm{Sp}(2) \cong \mathrm{Spin}(5)$ (where $\mathrm{Spin}(5)$ is the double cover of $\mathrm{SO}(5)$). Although we will only study the $\mathrm{SO}(5)$ NL$\sigma$M numerically in this paper, it is highly beneficial to consider its large-$N$ generalization, namely the $\mathrm{Sp}(2N)$ Grassmannian NL$\sigma$M defined on the target space $\frac{\mathrm{Sp}(2N)}{\mathrm{Sp}(N)\times \mathrm{Sp}(N)}$. Interestingly, this $\mathrm{Sp}(2N)$ Grassmannian NL$\sigma$M can also be straightforwardly realized using LLL.

We begin with $4N$-flavor fermions $\psi_a$ ($a=1,\dots,4N$) in the LLL, possessing a maximal flavor symmetry of $\mathrm{SU}(4N)$ alongside $\mathrm{U}(1)$ charge conservation. Next we introduce interactions that break the $\mathrm{SU}(4N)$ symmetry down to $\mathrm{Sp}(2N)$ symmetry. The $4N$-flavor fermions $\hat{\mathbf{\Psi}}=(\psi_1\dots\psi_{4N})^\mathrm{T}$ form an $\mathrm{Sp}(2N)$ fundamental, resulting in $\hat{\mathbf{\Psi}}^\mathrm{T}\mathbf{J}\hat{\mathbf{\Psi}}$ being invariant under $\mathrm{Sp}(2N)$ but not $\mathrm{SU}(4N)$, where $\mathbf{J}=\left(\begin{matrix}0&\mathbb{I}_{2N}\\-\mathbb{I}_{2N}&0\end{matrix}\right)$. Consequently, we can consider a Hamiltonian in the LLL with a real space interaction,
\begin{multline}
    H_\textrm{int}=\int\mathrm{d}^2 \vec{r}_1 \,\mathrm{d}^2 \vec{r}_2\left[U(\vec{r}_{12}) \hat{n}(\vec{r}_1)\hat{n}(\vec{r}_2)\vphantom{\frac{V(\vec{r}_{12})}{2N}}\right.\\\left.-\frac{V(\vec{r}_{12})}{2N}\hat{\Delta}^\dagger(\vec{r}_1)\hat{\Delta}(\vec{r}_2)\right].
    \label{eq:2a_hmt}
\end{multline}
where $\hat{n}(\vec r)=\hat{\mathbf{\Psi}}^\dagger(\vec r)\hat{\mathbf{\Psi}}(\vec r)$ and $\hat{\Delta}(\vec r)=\hat{\mathbf{\Psi}}^\mathrm{T}(\vec r)\mathbf{J}\hat{\mathbf{\Psi}}(\vec r)$. For simplicity we consider the potentials to be both $\delta$-functions $U(\vec{r}_{12})=U\delta(\vec{r}_{12}),V(\vec{r}_{12})=V\delta(\vec{r}_{12})$. The first term can be viewed as a continuum version of $\mathrm{SU}(4N)$ Hubbard interaction on the lattice, which maintains the maximal $\mathrm{SU}(4N)$ fermion flavor symmetry. The second term breaks $\mathrm{SU}(4N)$ down to $\mathrm{Sp}(2N)$ symmetry. It is worth noting that when $N=1$ our model reduced to the $\mathrm{SO}(5)$ NL$\sigma$M studied in Refs.~\cite{Ippoliti2018Half,Wang2021Phases}, which is expressed in a slightly different form~\footnote{The Hamiltonian in Refs.~\cite{Ippoliti2018Half,Wang2021Phases}
\begin{equation*}
    H_\textrm{int}=\int\mathrm{d}^2 \vec{r}\left[\tilde{U}_0\hat{n}(\vec{r})^2-\tilde{U}(\hat{\mathbf{\Psi}}^\dagger\gamma^i\hat{\mathbf{\Psi}})(\vec{r})^2\right]
\end{equation*}
converts into our Hamiltonian as
\begin{align*}
    U(\vec{r}_1,\vec{r}_2)&=(\tilde{U_0}+3\tilde{U})\delta(\vec{r}_{12}),\\
    V(\vec{r}_1,\vec{r}_2)&=8\tilde{U}\delta(\vec{r}_{12}),
\end{align*}
where $\tilde{U}_0$ and $\tilde{U}$ are the denoted $U_0$ and $U$ in Refs.~\cite{Ippoliti2018Half,Wang2021Phases}}.

Let us now explain why Eq.~\eqref{eq:2a_hmt} on LLL at half filling gives an $\mathrm{Sp}(2N)$ Grassmannian NL$\sigma$M with a level-1 WZW term. The $N=1$ case has been discussed in Refs.~\cite{Lee2015Wess,Ippoliti2018Half}.  When $V=0$, the dynamics of the system is captured by a NL$\sigma$M on the $\mathrm U(4N)$ Grassmannian $\frac{\mathrm U(4N)}{\mathrm U(2N)\times \mathrm U(2N)}$,
\begin{equation} \label{eq:NLSM}
S[\mathbf Q] = \frac{1}{g} \int \mathrm d^2 \vec r\,\mathrm d t\operatorname{Tr}(\partial^\mu\mathbf  Q(\vec r, t))^2 + S_\mathrm{WZW}[\mathbf Q].
\end{equation}
Here $\mathbf Q(\vec r, t)$ is a $4N\times 4N$ matrix field living on the $\mathrm U(4N)$ Grassmannian, parameterized by
\begin{equation}
    \mathbf Q = \mathbf A^\dagger \left(\begin{matrix}
        \mathbb{I}_{2N} & 0 \\ 0 & - \mathbb{I}_{2N}
    \end{matrix}\right) \mathbf A,
\end{equation}
with $A$ being a $\mathrm U(4N)$ matrix. The matrix field $\mathbf Q(\vec r, t)$ encodes the occupation of fermions in our original system, specifically describing which $2N$ fermions out of the total $4N$ are occupied at the space-time coordinate $(\vec r, t)$. The same theory has also been proposed to describe the surface of certain $(3+1)$D symmetry protected topological phase~\cite{Xu2013NLSM}.  The WZW term has a simple physical interpretation: the skyrmion, characterized by $\pi_2\left(\frac{\mathrm U(4N)}{\mathrm U(2N)\times\mathrm U(2N)}\right) = \mathbb Z$, is a fermion carrying a $\mathrm{U}(1)$ electronic charge~\cite{Komargodski2018}. This generalizes a well-established result of the quantum Hall ferromagnet~\cite{Sondhi1993}, which corresponds to the case of $N=1/2$ in our scenario. Specifically, one can consider a special skyrmion that exhibits non-trivial patterns solely in the first two components of fermions, which then reduces to the familiar story of the quantum Hall ferromagnet.

Once a finite $V$ is introduced, the global $\mathrm{SU}(4N)$ symmetry is explicitly broken down to the $\mathrm{Sp}(2N)$ symmetry. As a consequence, the matrix field $\mathbf Q(\vec r, t)$ becomes energetically favorable to fluctuate on the $\mathrm{Sp}(2N)$ Grassmannian $\frac{\mathrm{Sp}(2N)}{\mathrm{Sp}(N)\times \mathrm{Sp}(N)}$, which is a sub-manifold of the larger $\mathrm U(4N)$ Grassmannian. Additionally, the WZW term defined on the $\mathrm U(4N)$ Grassmannian is also reduced to a WZW term on the $\mathrm{Sp}(2N)$ Grassmannian. Therefore, even at finite $V$, the system can still be effectively described by Eq.~\eqref{eq:NLSM}, where the matrix field $\mathbf Q(\vec{r}, t)$ resides on the $\mathrm{Sp}(2N)$ Grassmannian and is parameterized by
\begin{equation}
    \mathbf Q = \mathbf M^\mathrm{T}\begin{pmatrix}
        \mathbf{J}_{N} & 0 \\ 0 & -\mathbf{J}_{N}
    \end{pmatrix}\mathbf M, \quad \mathbf M\in \mathrm{Sp}(2N).
\end{equation}
The stiffness of the NL$\sigma$M is controlled by $V/U$ in our original model given by Eq.~\eqref{eq:2a_hmt}. The specific value of $g$ will determine the phase of the system, which we will discuss in further detail later. 

\begin{figure}[b]
    \centering
    \includegraphics[width=\linewidth]{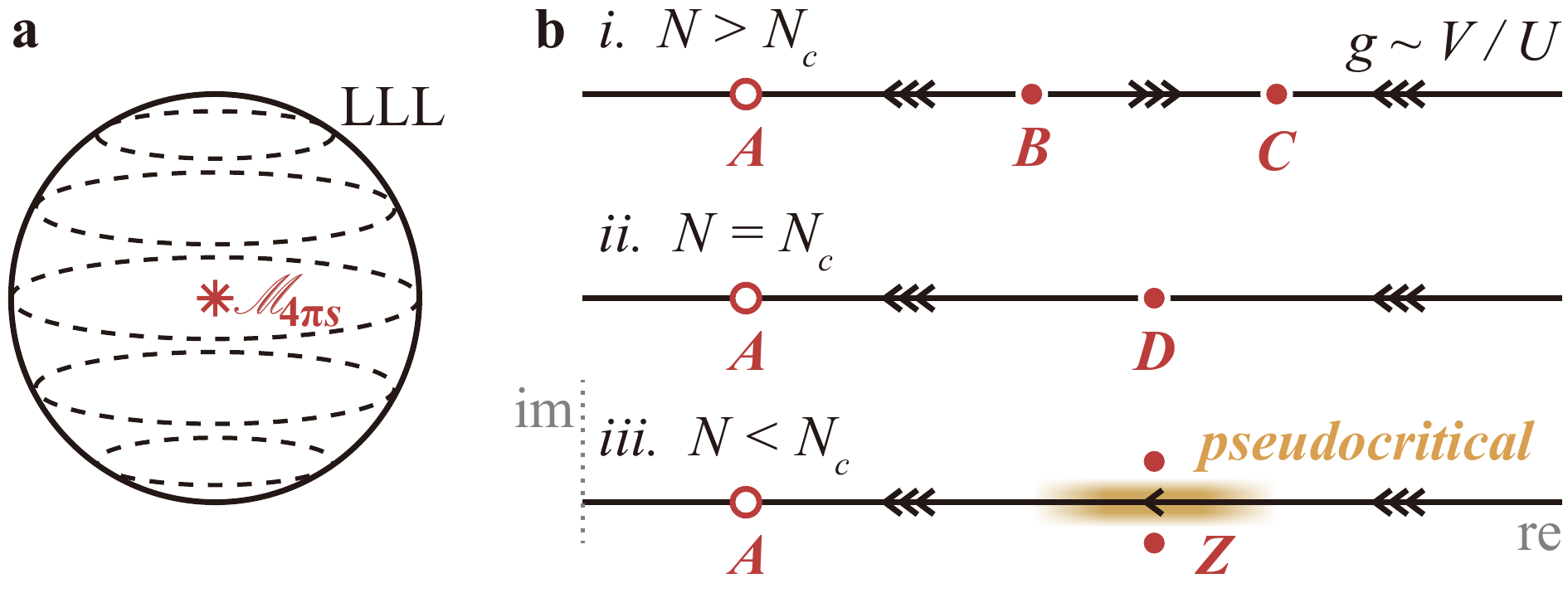}
    \caption{(a) An illustration of the fuzzy sphere setup. (b) Possible scenarios of the RG flow of the $\mathrm{Sp}(N)$ model as a function of $V/U$: (i) At $N>N_c$, there is a symmetry-breaking fixed point $A$, a repulsive $B$ and an attractive $C$ CFT fixed points; (ii) at $N=N_c$, the CFT fixed points $B$ and $C$ merges into a single fixed point $D$; (iii) at $N<N_c$, $D$ splits into two complex fixed points $Z$, and the region in vicinity exhibits pseudo-critical behavior. Here the filled and empty circles denote CFT and non-CFT fixed points.}
    \label{fig:1}
\end{figure}

Having understood how to realize the $\mathrm{Sp}(2N)$ Grassmannian NL$\sigma$M on the LLL, we are now ready to extend it to the sphere. In practice, we simply consider the LLL on a sphere with a $4\pi s$ monopole placed at the center~\cite{Haldane1983Fractional} (Fig.~\ref{fig:1}a). The LLL on the sphere consists of $N_\mathrm{orb} = 2s+1$ degenerate orbitals, which can be described by the monopole harmonics $Y_{sm}^{(s)}(\theta,\varphi)$~\cite{Wu1976Dirac}, where $m=-s, -s+1, \cdots, s$. These $2s+1$ orbitals form a spin-$s$ irreducible representation of the $\mathrm{SO}(3)$ sphere rotation. On the sphere, we can parameterize the system using spherical coordinates $(\theta, \varphi)$, and we have $\mathrm{d}^2 \vec r = N_\mathrm{orb} \sin\theta\,\mathrm{d}\theta\,\mathrm{d}\varphi$, and $\delta(\vec r_{12}) = \frac{1}{N_\mathrm{orb}} \delta(\cos\theta_1 - \cos\theta_2) \delta(\varphi_1 - \varphi_2)$. Here, we utilize the fact that on the spherical LLL, the sphere radius $R$ and the Landau orbital number $N_\mathrm{orb}$ are physically equivalent, i.e., $N_\mathrm{orb} \sim R^2$.

Next we project the interaction Eq.~\eqref{eq:2a_hmt} onto the LLL on sphere,
\begin{equation}
    \hat{\mathbf{\Psi}}(\theta,\varphi)=\frac{1}{\sqrt{N_\mathrm{orb}}}\sum_{m=-s}^s\bar{Y}_{sm}^{(s)}(\theta,\varphi)\hat{\mathbf{c}}_{m}.
\end{equation}
Here $\hat{\mathbf{c}}_{m}=(c_{m,1}, \cdots, c_{m,4N})$ is the second quantized fermion operator defined on the Landau orbitals. The normalization factor $1/\sqrt{N_\mathrm{orb}}$ is to ensure the particle density $\hat{n}(\vec r)=\hat{\mathbf{\Psi}}^\dagger(\vec r)\hat{\mathbf{\Psi}}(\vec r)$ is an intensive quantity. Rewriting the interaction Hamiltonian in a second quantized form in terms of $\hat{\mathbf{c}}_{m}$, we get
\begin{widetext}
\begin{equation}
    H_\textrm{int}^\textrm{(LLL)}=\sum_{m_1m_2m_3m_4}\left(U_{m_1m_2m_3m_4}(\hat{\mathbf{c}}^\dagger_{m_1}\hat{\mathbf{c}}_{m_4})(\hat{\mathbf{c}}^\dagger_{m_2}\hat{\mathbf{c}}_{m_3})-V_{m_1m_2m_3m_4}(\hat{\mathbf{c}}^\dagger_{m_1}\mathbf{J}^\dagger\hat{\mathbf{c}}^\dagger_{m_2})(\hat{\mathbf{c}}_{m_3}\mathbf{J}\hat{\mathbf{c}}_{m_4})\right).
\end{equation}
The parameter $U_{m_1m_2m_3m_4}$ is connected to the Haldane pseudopotential $U_l$ by
\begin{equation}
    U_{m_1m_2m_3m_4}=\sum_lU_l(4s-2l+1)\begin{pmatrix}
        s&s&2s-l\\m_1&m_2&-m_1-m_2
    \end{pmatrix}\begin{pmatrix}
        s&s&2s-l\\m_4&m_3&-m_4-m_3
    \end{pmatrix}\delta_{m_1+m_2,m_3+m_4},
\end{equation}
\end{widetext}
where $\begin{pmatrix}
    j_1&j_2&j_3\\m_1&m_2&m_3
\end{pmatrix}$
is the Wigner $3j$ symbol, and similarly $V_{m_1m_2m_3m_4}$ is connected to the Haldane pseudopotential $V_l$. For the case of $\delta$ potential we choose, the only non-zero component of the Haldane pseudopotential is $l=0$.

At last, let us comment on why the LLL projection (truncation) leads to a fuzzy sphere. It is instructive to consider the projection of the coordinates of a unit sphere, denoted as $\vec x = (\sin\theta\cos\varphi, \sin\theta\sin\varphi, \cos\theta)$. After the projection, the coordinates are transformed into $(2s+1)\times (2s+1)$ matrices, where $(\vec X)_{m_1,m_2}= \int \sin\theta\,\mathrm{d}\theta\,\mathrm{d}\varphi \, \vec x\, \bar Y^{(s)}_{s, m_1}(\theta,\varphi) Y^{(s)}_{s, m_2}(\theta,\varphi)$. These matrices satisfy the following relations:
\begin{equation}
    [\mathbf{X}^\mu, \mathbf{X}^\nu] = \frac{1}{s+1} i\epsilon^{\mu\nu\rho}\mathbf{X}_\rho, \quad \mathbf{X}_\mu\mathbf{X}^\mu = \frac{s}{s+1} \mathbf 1_{2s+1}.
\end{equation}
The fact that the three coordinates satisfy the $\mathrm{SO}(3)$ algebra formally defines a fuzzy sphere \cite{Madore1992Fuzzy}. It is worth noting that in the thermodynamic limit $s\rightarrow\infty$, the fuzziness disappears and a unit sphere is recovered.

\subsection{Conformal window and pseudo-criticality}
\label{sec:2b}

As conjectured in Ref.~\cite{Zou2021Stiefel}, the generic phase diagram of a 3D NL$\sigma$M with a WZW term can be summarized in Fig.~\ref{fig:1}b, which contains three different situations, depending on the value of $N$:
\begin{enumerate}[label=\textit{\roman*}.]
    \item $N>N_c$. For $N$ larger than a critical value $N_c$, there exist three different fixed points as we tune the NL$\sigma$M coupling $g$, or equivalently $V/U$ in our model. When $g$ is small, the NL$\sigma$M spontaneously breaks the global symmetry, and the ground state manifold corresponds to the target space of the NL$\sigma$M (in our case, the $\mathrm{Sp}(2N)$ Grassmannian). This corresponds to the attractive fixed point $A$.
    On the other hand, when $g$ is large, the system flows to the attractive fixed point C, which is fully symmetric and described by a 3D CFT. For the $\mathrm{Sp}(2N)$ Grassmannian NL$\sigma$M considered in our paper, this attractive conformal fixed point is in the same universality class as a QCD$_3$ theory with $2N$ flavors of Dirac fermions coupled to an $\mathrm{SU}(2)$ gauge field~\cite{Komargodski2018,Zou2021Stiefel}. In the limit of $N\gg 1$, the operator spectrum can be computed using the standard large-$N$ expansion technique~\cite{Xu2008Renormalizaiton}. This QCD$_3$ theory has also been studied by lattice Monte Carlo simulation~\cite{Karthik2018QCD3}.
    At a critical coupling $g_c$, there is a repulsive fixed point $B$, which also corresponds to a fully symmetric CFT. This fixed point describes the phase transition between the CFT phase ($C$) and the spontaneous symmetry-breaking phase ($A$).
    \item $N=N_c$. When decreasing $N$ to a critical value $N=N_c$, the two CFT fixed points $B$ and $C$ merge into one ($D$) and the singlet $S$ becomes exactly marginal ($\Delta_S=3$).
    \item $N<N_c$. For $N<N_c$, the fixed point $D$ splits into two fixed points located in the complex plane, denoted $Z$. Along the real axis, there are no CFT fixed points. However, the complex fixed points $Z$ are described by complex CFTs, which have complex conformal data including complex scaling dimensions~\cite{Gorbenko2018Walking,Gorbenko2018Walking2}. When $Z$ is sufficiently close to the real axis, these complex conformal data have a very small imaginary part. Importantly, the RG flow near $Z$ is slow, and over a large length scale, the system exhibits an approximate conformal symmetry with conformal data that closely resembles the real part of the complex CFT's complex conformal data. This behavior, referred to as pseudo-criticality or ``walking behavior'', is conjectured to account for the anomalous scaling observed numerically in the DQCP. A similar phenomenon has been observed in the 2D 5-state Potts model~\cite{Ma2019Potts}.
\end{enumerate}

So the key question pertains to the value of $N_c$ for the $\mathrm{Sp}(2N)$ Grassmannian NL$\sigma$M. If $N_c<1$, the $\mathrm{SO}(5)$ DQCP corresponds to a genuine continuous transition described by the attractive conformal fixed point $C$ shown in Fig.\ref{fig:1}b, i. On the other hand, if $N_c>1$, the $\mathrm{SO}(5)$ DQCP exhibits pseudo-critical behavior as depicted in Fig.\ref{fig:1}b, iii.
We also note that such $N$-dependent phase diagrams are believed to be common in various models and theories. For instance, in critical gauge theories involving Dirac fermions (or critical bosons) coupled to a dynamical gauge field $G(k)=\mathrm{SU}(k)$, $\mathrm{U}(k)$, $\mathrm{Sp}(k)$, etc., there exists a critical value $N_c(G(k))$ for each gauge group~\cite{Kaplan2009Conformality}. Our $\mathrm{Sp}(2N)$ Grassmannian NL$\sigma$M corresponds to $2N$ Dirac fermions coupled to an $\mathrm{SU}(2)$ gauge field.  However, determining the precise region of the conformal window ($N>N_c(G(k))$) for any gauge theory has been a longstanding challenge in the field. The main difficulty lies in distinguishing between pseudo-critical behavior and true critical behavior: the former also exhibits an approximate conformal symmetry over a large length scale, while the conformal symmetry of the latter is exact only in the thermodynamic limit.

Here, we propose that the fuzzy sphere microscope can be used to resolve the outstanding puzzle of the conformal window. The idea is to examine the RG flow of operators' scaling dimensions, particularly, the lowest-lying global symmetry singlet $S$ which controls RG flow. A characteristic feature of pseudo-criticality is that, at the coupling $V/U$ on the right-hand side of the vicinity of the complex fixed points $Z$, since the RG flow is attractive at the beginning towards the complex fixed points and then become repulsive past the complex fixed points, $\Delta_S$ will decrease from being slightly irrelevant (i.e., $\Delta_S \gtrsim 3$) to slightly relevant (i.e., $\Delta_S \lesssim 3$) as the system size increases~\footnote{Another reason that $S$ eventually becomes relevant is that the complex fixed points themselves have $\operatorname{Re}(\Delta_S)\lesssim3$.}. Such flow will not occur in the case of $N>N_c$, as one will have either $\Delta_S > 3$ or $\Delta_S$ increasing from being relevant to irrelevant as the system size increases. The latter situation corresponds to the coupling being close to the critical coupling of the repulsive fixed point $B$. A quantitative analysis based on conformal perturbation will follow in the next section.
In this paper, we will focus on the case $N=1$, where we find a clear signature of pseudo-criticality.

\subsection{Conformal perturbation for pseudo-criticality}
\label{sec:2b_1}

The RG flow can be quantitatively described by the conformal perturbation theory~\cite{Gorbenko2018Walking,Gorbenko2018Walking2}. In the scenario of psudocriticality, we can write the Hamiltonian as 
\begin{equation}
    H(\lambda)=H_0+\lambda\int\frac{\mathrm{d}\vec{r}}{4\pi}S(\vec{r})
\end{equation}
where $H_0$ is the Hamiltonian at a reference point, and $\lambda=\lambda(R,\lambda_0)$ is the factor of the singlet operator $S$ that depend on the linear system size $R=N_\mathrm{orb}^{1/2}$ and a tuning parameter $\lambda_0$ in the Hamilonian that is determined by $V/U$. The rescaled energy of an arbitrary operator $\Phi$ could be interpreted as the scaling dimension due to the state-operator correspondence in the presence of conformal symmetry
\begin{equation}
    \label{eq:dim_perturb}
    \Delta_\Phi(\lambda)=\langle\Phi|H(\lambda)|\Phi\rangle+\mathscr{O}(\lambda^2)=\Delta_{\Phi,0}+\lambda f_{\Phi\Phi S}+\mathscr{O}(\lambda^2)
\end{equation}
to the lowest order of $\lambda$, where $\Delta_{\Phi,0}=\langle\Phi|H_0|\Phi\rangle$~\footnote{The calibration is done by rescaling the stress tensor $\Delta_{\mathscr{T}^{\mu\nu}}=3$ and $\Delta_{J^\mu}=2$. The rescaling factor is not affected by the perturbation $S$ because of the im-reflection property $f_{JJS}\propto\operatorname{im}\Delta_J=0$.}.

The flow of $\lambda(R,\lambda_0)$ is captured by the $\beta$-function 
\begin{equation}
    \beta(\lambda)=R\frac{\mathrm{d}\lambda(R,\lambda_0)}{\mathrm{d}R}
\end{equation}
To capture the structure of the the pseudo-critical flow, to the lowest order the $\beta$-function takes the form
\begin{equation}
    \beta(\lambda)=R\frac{\mathrm{d}\lambda(R,\lambda_0)}{\mathrm{d}R}=-\alpha(\lambda^2+y^2)+\mathscr{O}(y^4).
\end{equation}
where $\lambda=0$ is defined to be the point where the flow is the slowest, and $\alpha$ and $y$ are parameters to be determined. Continuating $\lambda$ into the complex plane, the two complex fixed points lie at $\lambda^*_{Z_\pm}=\pm iy$~\footnote{Note that the notation is different from Refs.~\cite{Gorbenko2018Walking,Gorbenko2018Walking2}, $g^*=0,-2iy$ and $g=\lambda-iy$.}. Subsequently, the scaling dimensions at two complex fixed points $\Delta_{\Phi,Z_\pm}$ are given by 
\begin{align}
    \Delta_{\Phi,Z_\pm}&=\Delta_\Phi(\pm iy)=\Delta_{\Phi,0}\pm iyf_{\Phi\Phi S}+\mathscr{O}(y^2).
\end{align}

The physical parameters lie on the real axis $\lambda\in\mathbb{R}$. One can solve from the flow equation that to the leading order
\begin{equation}
    \lambda(R,\lambda_0)=y\tan\left[\tan^{-1}\frac{\lambda_0}{y}-\alpha y\log\frac{R}{R_0}\right]+\mathscr{O}(y^2)
    \label{eq:running_coupling}
\end{equation}
The rescaled energy of any state corresponding to the primary operator $\Phi$ is given by substituting into Eq.~\eqref{eq:dim_perturb}. To the leading order
\begin{equation}
    \Delta_\Phi(\lambda_0,R)=\Delta_{\Phi,0}+f_{\Phi\Phi S}\lambda(R,\lambda_0)+\mathscr{O}(\lambda^2).
    \label{eq:pscr_scal}
\end{equation}

Physically, we can compare it with the scenario where there are two real conformal fixed points, which are very close to each other. We set the repulsive fixed point at $\lambda=-x$ and the attractive fixed point at $\lambda=+x$. To the leading order,
\begin{align}
    \tilde\beta(\lambda)&=-\alpha(\lambda^2-x^2)+\mathscr{O}(x^4)\nonumber\\
    \tilde\lambda(R,\lambda_0)&=x\tanh\left[\tanh^{-1}\frac{\lambda_0}{x}+\alpha x\log\frac{R}{R_0}\right]+\mathscr{O}(x^2)\nonumber\\
    \tilde\Delta_\Phi(\lambda_0,R)&=\Delta_\Phi+\tilde\lambda f_{\Phi\Phi S}+\mathscr{O}(\tilde\lambda^2)
\end{align}
where we use the tilde to distinguish with the case of pseudo-criticality, and we define $\tanh^{-1}\xi=i\pi-\coth^{-1}\xi$ when $|\xi|>1$. A pronounced difference is that $\Delta_\Phi$ increases with system size $R$ on one side of the attractive fixed point $\lambda^*=x$, and decreases with $R$ on the other side, and $\Delta_\Phi(\lambda)$ for different $R$ should intersect near the fixed point. A similar intersection is also expected in the vicinity of the repulsive fixed point. In contrast, for the case of pseudo-criticality, the dependence of $\Delta_\Phi$ on $R$ stays the same regardless of $\lambda$. This can be used to distinguish the two scenarios. We also note that the microscopic Hamiltonian may also contain various irrelevant operators $S',S''$, etc. Each of them has a subleading contribution to the scaling dimension $\delta\Delta_\Phi(R;\lambda'_0,\dots)\sim \lambda'_0R^{\Delta_{S'}-3}+\dots$ which will scale to zero in the thermodynamic limit. 

\begin{figure}[b!]
    \centering
    \includegraphics[width=\linewidth]{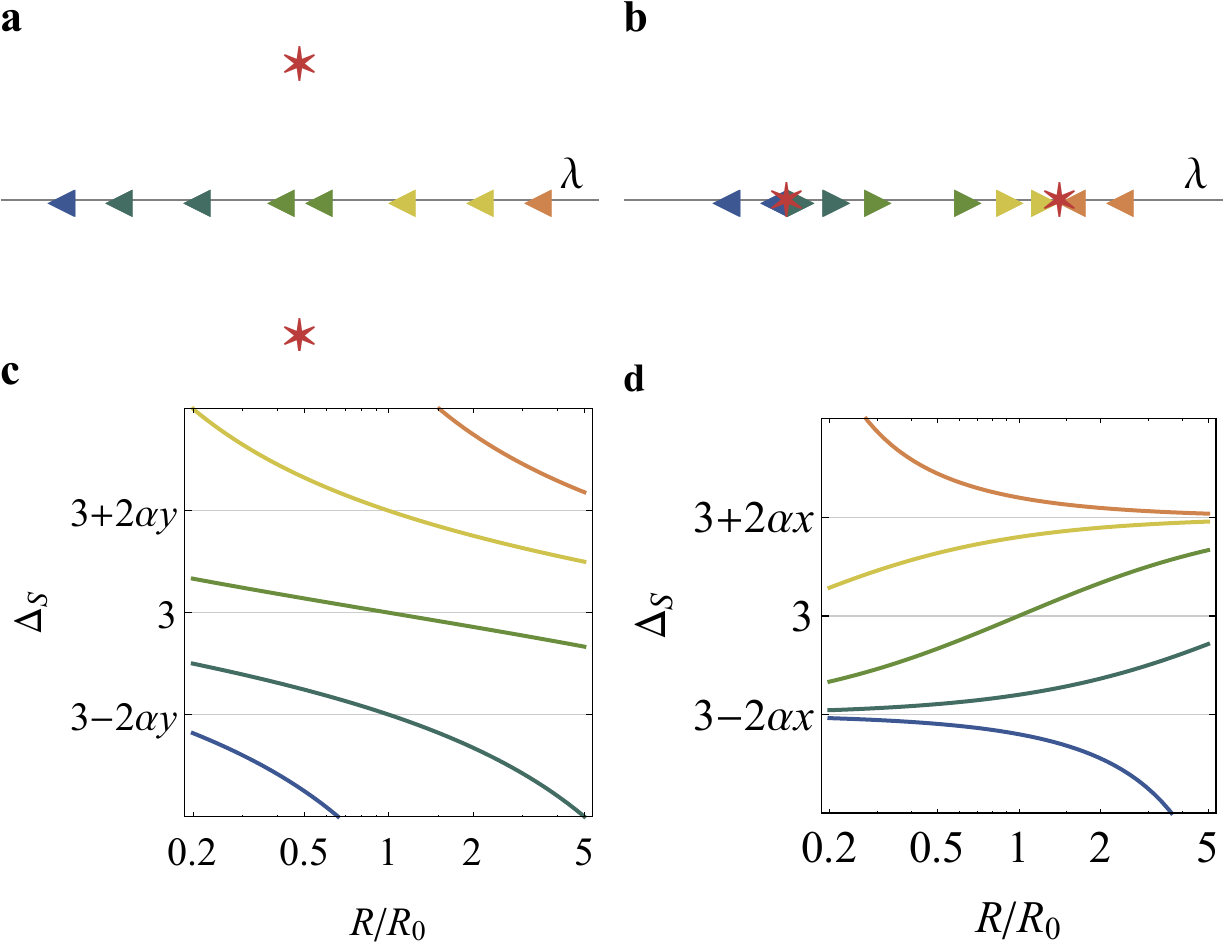}
    \caption{An illustration of the flow diagram of $\lambda$ (a,b) and the dependence of the rescaled energy $\Delta_S(\lambda,R)$ on the system size $R$ for different parameter $\lambda$ (c,d) in the scenario of pseudo-criticality (a,c) and true criticality with two real fixed points (b,d). For the purpose of illustration, we set $\alpha y=0.2$ and $\alpha x=0.5$ and take the horizontal axes in logarithmic scale. The arrows in (a,b) indicate the position in the flow at scale $R/R_0=0.5$ and $2$ for different curves in (c,d). The red stars denote the fixed points. A characteristic feature of pseudo-criticality (a,c) is that the lowest singlet will always flow from irrelevant to relevant as the system size $R$ increases, if the flow starts from $\Delta_S>3$. }
    \label{fig:1_1}
\end{figure}

This result also supports our claim that the RG flow of the lowest singlet from irrelevant to relevant is a key feature for pseudo-criticality. To the leading order $\Delta_S=3+\mathscr{O}(y^2)$ and $f_{SSS}=2\alpha+\mathscr{O}(y^2)$. Thus, from Eq.~\eqref{eq:pscr_scal}, the rescaled energy for the state corresponding to $S$
\begin{equation}
    \Delta_S(\lambda_0,R)=3-2\alpha y\tan\left[\alpha y\log\frac{R}{R_c(\lambda_0)}\right]+\mathscr{O}(y^2)
\end{equation}
where $R_c(\lambda_0)=R_0\exp\frac{\tan^{-1}\lambda_0/x}{\alpha x}$. For small system size $R<R_c(\lambda_0)$, $S$ is irrelevant $\Delta_S(\lambda_0,R)>3$ ; as $R$ increases, $\Delta_S$ decreases, and until $R>R_c$, $S$ will become relevant $\Delta_S(\lambda_0,R)<3$. The size dependence of $\Delta_S$ is sketched in Fig.~\ref{fig:1_1}c We can similarly work out the expression in the case where there are two real fixed points,
\begin{align}
    \tilde\Delta_S(\lambda_0,R)&=3+2\alpha x\tanh\left[\alpha x\log\frac{R}{\tilde R_c(\lambda_0)}\right]+\mathscr{O}(x^2)
\end{align}
where $\tilde R_c(\lambda_0)=R_0\exp\frac{\tanh^{-1}\lambda_0/x}{\alpha x}$, and $\Delta_S(\lambda_0,R)$ increases from relevant $\Delta_S(\lambda_0,R<R_c)<3$ to irrelevant $\Delta_S(\lambda_0,R>R_c)>3$ along the flow. The size dependence of $\Delta_S$ is sketched in Fig.~\ref{fig:1_1}d.
We also note that in the vicinity of the fixed points, the scale $\tilde{R_c}\to 0$ as $\lambda_0\to-x$ and $\tilde{R_c}\to \infty$ as $\lambda_0\to+x$, and we can use the asymptotic behavior $\tanh\xi\to\pm(1-2e^{\mp\xi})$ as $\xi\to\pm\infty$ to recover the familiar flow behavior
\begin{multline}
    \tilde\Delta_S(\lambda_0,R)\\\to\left\{\begin{array}{ll}
        3+2\alpha x+2\alpha(\lambda_0-x)(R/R_0)^{-2\alpha x},&\lambda_0\to+x\\
        3-2\alpha x+2\alpha(\lambda_0+x)(R/R_0)^{+2\alpha x},&\lambda_0\to-x\\
    \end{array}\right..
\end{multline}

In conclusion, the conformal perturbation gives the dependence of the rescaled energy for any primary operator $\Phi$ from state-operator correspondence on linear system size $R$ and parameter $\lambda_0$ in Hamiltonian
\begin{multline}
    \Delta_\Phi(\lambda_0,R)=\Delta_{\Phi,0}\\+f_{\Phi\Phi S}y\tan\left[\tan^{-1}\frac{\lambda_0}{y}-\alpha y\log\frac{R}{R_0}\right]+\mathscr{O}(y^2).
    \label{eq:perturb_final}    
\end{multline}
to the leading order. We will verify it numerically as evidence for pseudo-criticality in Sec.~\ref{sec:3d_1}.

\subsection{Relation to the original story of DQCP}
\label{sec:2c}
Before presenting our results, it is worth commenting on how our model is related to the original story of DQCP.
DQCP was originally proposed to describe a direct continuous transition between a N\'eel phase and VBS phase on the square lattice~\cite{Senthil2004Deconfined,Senthil2004Quantum}. The effective field theory is the CP$^1$ model, which is a gauge theory that has $N_f=2$ flavors of complex critical bosons coupled to a $\mathrm{U}(1)$ gauge field. This field theory has an explicit $\mathrm{SU}(2)\times \mathrm{U}(1)$ global symmetry, where $\mathrm{SU}(2)$ is the flavor rotation symmetry between the two flavors of bosons, and $\mathrm{U}(1)$ is also called the topological $\mathrm{U}(1)$ corresponding to the flux conservation of the $\mathrm{U}(1)$ gauge field. For the N\'eel-VBS transition on the square lattice, there is only an $\mathrm{SU}(2)\times \mathbb Z_4$ symmetry in the UV, where $\mathrm{SU}(2)$ is the spin-rotation symmetry, while $\mathbb Z_4$ is the square lattice $C_4$ rotation symmetry. At the phase transition, it is conjectured that $\mathbb Z_4$ is enhanced to $\mathrm{U}(1)$, which means that the $8\pi$-monopole $\mathscr M_{8\pi}$ has to be irrelevant. Similarly, for the N\'eel-VBS transition on the honeycomb lattice~\cite{Sumiran2013Neel,Ye2022Topological}, where there is only a $C_3$ lattice rotation, the $6\pi$-monopole $\mathscr M_{6\pi}$ has to be irrelevant if it is described by the DQCP.

More recently, it was numerically discovered that the $\mathrm{SU}(2)\times \mathrm{U}(1)$ symmetry enhances to the $\mathrm{SO}(5)$ symmetry~\cite{Nahum2015Emergent}, and it inspired a number of new dual descriptions of DQCP~\cite{Wang2017Deconfined}, including the $\mathrm{SO}(5)$ NL$\sigma$M studied here. In the original N\'eel-VBS transition, the $\mathrm{SO}(5)$ symmetry corresponds to the symmetry between the 3-component N\'eel order parameter and the 2-component VBS order parameter. The WZW term encodes the physics of intertwinement between the N\'eel and VBS orders, namely, the topological defect of one binds the symmetry charge of the other~\cite{Levin2004Deconfined,Tanaka2005Many,Senthil2006Competing}. One component of the $\mathrm{SO}(5)$ rank-2 symmetric traceless tensor becomes $\mathrm{SU}(2)\times \mathrm{U}(1)$ singlet, so it is the tuning operator for the N\'eel-VBS transition. In our model, this operator is not allowed by the $\mathrm{SO}(5)$ symmetry, so there is no relevant singlet if the DQCP is a genuine critical point without further fine-tuning.

It is also worth noting that the DQCP exhibits a mixed anomaly between the $\mathrm{SO}(5)$ and time-reversal symmetry~\cite{Wang2017Deconfined}. In the N\'eel-VBS transition, apart from time-reversal symmetry, there is only an $\mathrm{SU}(2)\times \mathbb Z_4$ symmetry in the UV, which is consistent with this anomaly. Interestingly, the $\mathrm{SO}(5)$ NL$\sigma$M possesses exact $\mathrm{SO}(5)$ and time-reversal symmetry (i.e., particle-hole symmetry) in the UV, which appears to contradict the anomaly of DQCP. The way to reconcile this contradiction is by understanding that the particle-hole symmetry on the LLL is a non-local symmetry~\cite{Ma2022Edge}. Similar physics also applies to the $\mathrm{Sp}(2N)$ Grassmannian NL$\sigma$M, which realizes $2N$ flavor $\mathrm{SU}(2)$ QCD$_3$ with exact $\mathrm{Sp}(2N)$ and time-reversal symmetry in the UV. In contrast, for a lattice realization of QCD$_3$, one can only have $\mathrm{Sp}(N)\times \mathrm{Sp}(N)$ symmetry in the UV due to the parity anomaly.

\begin{figure*}[t!]
    \centering
    \includegraphics[width=\linewidth]{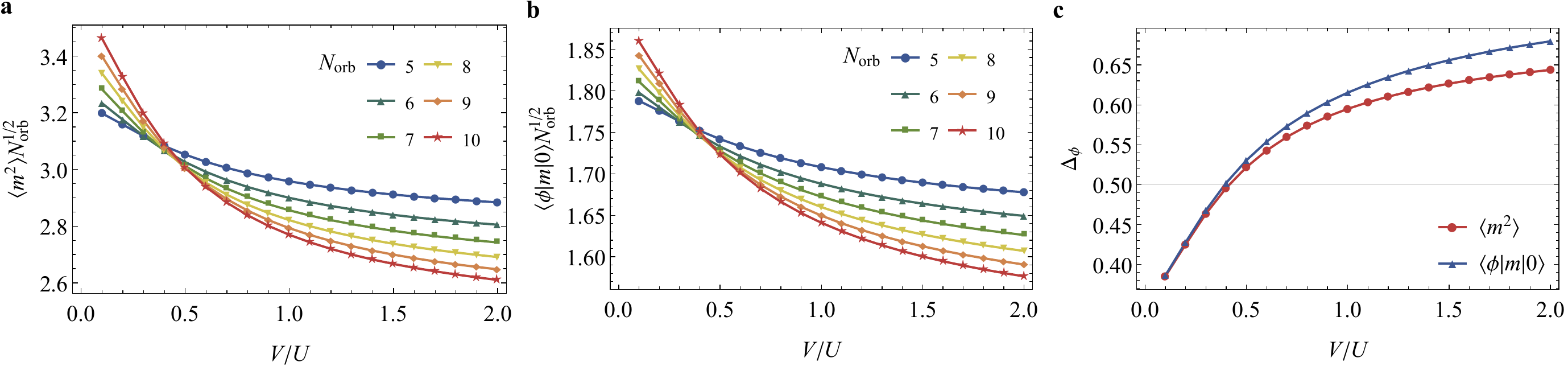}
    \caption{Identifying a disorder region by finite-size scaling of the order parameter $\mathrm{SO}(5)$ vector $m$. (a) $\langle m^2\rangle N_\mathrm{orb}^{1/2}$ and (b) $\langle\phi|m|0\rangle N_\mathrm{orb}^{1/4}$ as a function of $V/U$ at $5\leq N_\mathrm{orb}\leq 10$. (c) The scaling dimension $\Delta_\phi$ as a function of $V/U$ extracted from the finite-size scaling of $\langle m^2\rangle$ and $\langle\phi|m|0\rangle$. }
    \label{fig:2}
\end{figure*}

\section{Numerical results}
\label{sec:3}

\subsection{Exact diagonalization and quantum numbers}
\label{sec:3a}

We perform an exact diagonalization calculation for the Hamiltonian Eq.~(\ref{eq:2a_hmt}) at $N=1$ to get the lower spectra up to a system size $N_\mathrm{orb}=10$ (i.e., $20$ electrons). As each eigenstate carries a definite quantum number of all the symmetries of the Hamiltonian, and implementing the corresponding conserved quantities can divide the Hilbert space into sectors and block-diagonalize the Hamiltonian, it is useful to first analyze the symmetries of the Hamiltonian:

\begin{enumerate}[label=(\arabic*)]
    \item The $\mathrm{SO}(3)$ rotation symmetry of $S^2$. The angular momentum of the state can be determined by measuring the $\mathrm{SO}(3)$ quadratic Casimir $\langle\Phi|\hat{L}^2|\Phi\rangle=\ell_\Phi(\ell_\Phi+1)$. In the calculation, we implement the conservation of $\hat{L}^z$ which gives each state a quantum number $m^z_\Phi=\langle\Phi|\hat{L}^z|\Phi\rangle$. The $m^z=0$ sector can be further divided into even-$\ell$ and odd-$\ell$ sectors by representation under $\pi$-rotation around $y$-axis $\mathscr{R}_y$ (cf. Appendix \ref{sec:app_qn}).
    \item The $\mathrm{SO}(5)=\mathrm{Sp}(2)/\mathbb{Z}_2$ flavor symmetry. The $\mathrm{SO}(5)$ representation carried by the operator (state) is determined by the counting degeneracy of the corresponding states and by measuring the $\mathrm{SO}(5)$ quadratic Casimir. In this paper, we label the representations by their dimensions, namely, the singlet representation $\mathbf{1}$, the $\mathrm{SO}(5)$ vector representation (i.e., the $\mathrm{Sp}(2)$ antisymmetric rank-$2$ traceless tensor representation) $\mathbf{5}$, the $\mathrm{SO}(5)$ antisymmetric rank-$2$ tensor representation (i.e., the $\mathrm{Sp}(2)$ symmetric rank-$2$ tensor representation) $\mathbf{10}$, the $\mathrm{SO}(5)$ symmetric rank-$2$ traceless tensor representation $\mathbf{14}$. In the calculation, we implement two commuting conserved quantities $\sigma_{1,2}$ (cf. Appendix \ref{sec:app_qn}) due to this symmetry, corresponding to the two generators of the Cartan subalgebra of $\mathfrak{so}(5)$. The $\sigma_{1,2}=0$ sector can be further divided by representation under two permutations of layers $\mathscr{X}_{1,2}$ (cf. Appendix \ref{sec:app_qn}).
    \item The particle-hole symmetry $\mathscr{P}:\hat{\mathbf{c}}_m\to \mathbf{J}(\hat{\mathbf{c}}^\dagger_m)^\mathrm{T},i\to-i$. This further divides the $m^z=0$ sector.
\end{enumerate}

With these conserved quantities $L^z,\sigma_{1,2},\mathscr{P},\mathscr{R}_y$ and $\mathscr{X}_{1,2}$ implemented, we are able to look at the $500$ lowest lying states in the sector $L^z=0$ and $\sigma_{1,2}=0$ at maximal size $N_\mathrm{orb}=10$. The $N_\mathrm{orb}=10$ calculation takes $96.7\,\mathrm{GB}$ of memory and $7075$ seconds (using 2.4Hz Intel(R) Xeon(R) Gold 6148 CPU with 40 cores) to produce the $50$ lowest eigenstates in the maximal sector $(m^z,\sigma_1,\sigma_2,\mathscr{P},\mathscr{R}_y,\mathscr{X}_1,\mathscr{X}_2)=(0,0,0,+,+,+,+)$.

\subsection{Phase diagram}
\label{sec:3b}

To study the physics of DQCP, we should first identify a region in the phase diagram that does not show spontaneous symmetry breaking at finite system size. In the scenario of true criticality, there is an ordered and spontaneously symmetry broken phase at small $V/U$ and a CFT phase at large $V/U$, and the phase transition is described by another CFT. In the scenario of pseudo-criticality, in the thermodynamic limit, there is only one phase which is ordered and spontaneously symmetry broken. However, at finite system size, the behavior is similar to a crossover between a disordered region and an ordered region, as the system at small $V/U$ will flow to the vicinity of the symmetry-broken fixed point, and the system at large $V/U$ will flow towards the pseudo-critical region before its eventual ordering in the thermodynamic limit. As the RG flow in the pseudo-critical region is very slow and the correlation length is very large, a finite-size system in this region exhibits disorder. 

So we look at the $\mathrm{SO}(5)$ symmetry breaking order parameter (i.e. $\mathrm{SO}(5)$ vector),
\begin{equation}
    m^i=\frac{1}{N_\mathrm{orb}}\sum_m\hat{\mathbf{c}}^\dagger_m\gamma^i\hat{\mathbf{c}}_m,
\end{equation}
in the finite size systems, where the $\gamma$-matrices are $\{\mathbb{I}\otimes\tau^x,\mathbb{I}\otimes\tau^z,\sigma^x\otimes\tau^y,\sigma^y\otimes\tau^y,\sigma^z\otimes\tau^y\}$. In a quantum system described by a unitary CFT, the order parameter should scale with the linear scale of the system as $\langle m^2\rangle\sim R^{-2\Delta_\phi}=N_\mathrm{orb}^{-\Delta_\phi}$, where $\phi$ is the lowest parity-odd scalar $\mathrm{SO}(5)$ vector operator in the CFT. We can similarly look at the one-point function $\langle\phi|m|0\rangle\sim R^{-\Delta_\phi}$, where $|\phi\rangle$ is the state corresponding to the $\phi$ operator in the CFT. As its scaling dimension is bounded by the unitarity bound $\Delta_\phi\ge d/2-1=1/2$~\cite{Poland2019Conformal}, $\langle m^2\rangle N_\mathrm{orb}^{1/2}$ and $\langle\phi|m|0\rangle N_\mathrm{orb}^{1/4}$ should be decreasing with $N_\mathrm{orb}$. Numerically, we observe these two quantities increase with $N_\mathrm{orb}$ at $V/U\lesssim 0.4$, corresponding to a symmetry-breaking region, and decrease with $N_\mathrm{orb}$ at $V/U\gtrsim 0.5$~(Fig.~\ref{fig:2}a,b). We also perform a finite-size scaling to fit the scaling dimension $\Delta_\phi$~(Fig.~\ref{fig:2}c). These results consistently indicate that $\mathrm{SO}(5)$ symmetry is not breaking at $V/U>0.5$ for the system sizes $N_\mathrm{orb}\le 10$. Consequently, we will focus on this region in the following.

\begin{figure*}[t!]
    \centering
    \includegraphics[width=\linewidth]{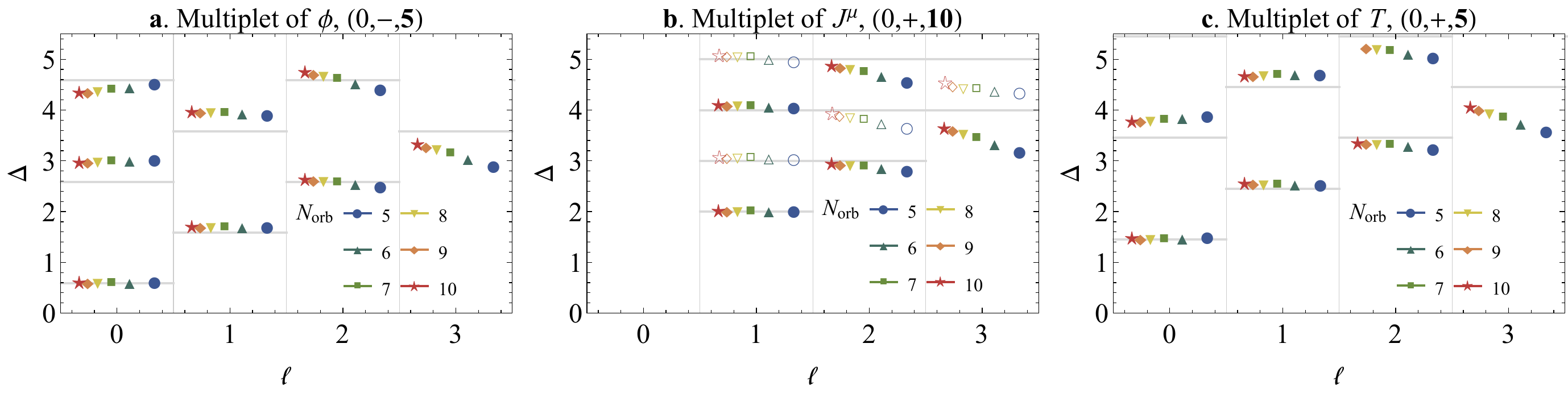}
    \caption{The scaling dimensions of the conformal multiplet of (a) the lowest $\mathrm{SO}(5)$ vector $\phi$, (b) the $\mathrm{SO}(5)$ symmetry current $J^\mu$ and (c) the lowest $\mathrm{SO}(5)$ rank-$2$ traceless symmetric tensor $T$ at different spin $\ell$ and system size $N_\mathrm{orb}$. The horizontal grey bar denotes the anticipated values based on the integer-spaced levels from the conformal symmetry. The filled and empty symbols in (b) signify the parity-even and parity-odd descendants. Parameter $V/U$ is taken such that $\Delta_\mathscr{T}=3$ exactly. The rest of the states in the respective sector, listed in Table~\ref{tbl:spec_scan}, can also be organized in conformal multiplets.}
    \label{fig:3}
\end{figure*}

\subsection{Approximate conformal symmetry}
\label{sec:3c}

To verify whether or not the $V/U\gtrsim 0.5$ region is described by a CFT, we need to examine if the energy spectrum has an emergent conformal symmetry, namely if they form irreducible representations of the conformal group. We first determine the size and parameter-dependent factor in Eq.~(\ref{eq:2c_state_operator}) by setting the scaling dimension of the $\mathrm{SO}(5)$ symmetry current to be exactly $\Delta_J=2$. The symmetries of the Hamiltonian should be identified with the symmetries of the conformal field theory, and so are the quantum numbers carried by the quantum states and by the CFT operators. In particular, the $\mathrm{SO}(3)$ rotation symmetry of $S^2$ is identified with the Lorentz rotation of the conformal group, and the angular momentum is identified with the Lorentz spin $\ell$; the particle-hole symmetry acts as an improper $\mathbb{Z}_2$ of $\mathrm{O}(3)$ and is thus identified as the spacetime parity $\mathscr{P}$ of the CFT~\cite{Zhu2023Uncovering}.

One convincing evidence for the conformal symmetry is the integer-spaced levels (e.g., see Ref.~\cite{Zhu2023Uncovering} for a detailed discussion). In a CFT spectrum, for any scalar primary $\Phi$ with quantum numbers $(\ell=0,\mathscr{P},\mathrm{rep.};\Delta)$, its descendants can be written in the form of $\partial^{\nu_1}\dots\partial^{\nu_j}(\partial^2)^n\Phi$ ($n,j\geq0$) with quantum numbers $(j,\mathscr{P},\mathrm{rep.};\Delta+2n+j)$; for spinning primary $\Phi^{\mu_1\dots\mu_\ell}$, its descendants can be written either as $\partial^{\nu_1}\dots\partial^{\nu_j}(\partial^2)^n\partial_{\mu_1}\dots\partial_{\mu_i}\Phi^{\mu_1\dots\mu_\ell}$ ($0\le i\le \ell, n,j\ge 0$) with quantum numbers $(\ell-i+j,\mathscr{P},\mathrm{rep.};\Delta+2n+i+j)$ or as $\epsilon_{\mu_1\rho\sigma}\partial^\rho\partial^{\nu_1}\dots\partial^{\nu_j}(\partial^2)^n\partial_{\mu_2}\dots\partial_{\mu_i}\Phi^{\mu_1\dots\mu_\ell}$ ($1\le i\le \ell, n,j\ge 0$) with quantum numbers $(\ell-i+j+1,-\mathscr{P},\mathrm{rep.};\Delta+2n+i+j)$; for conserved currents like $\mathrm{SO}(5)$ symmetry current $J^\mu$ and stress tensor $\mathscr{T}^{\mu\nu}$, only $i=0$ descendants exist due to the conservation $\partial_{\mu_1}\Phi^{\mu_1\dots\mu_\ell}=0$.

Numerically, we observe that the low-lying levels indeed exhibit a remarkable alignment with the integer-spaced patterns predicted by the conformal symmetry. Fig.~\ref{fig:3} shows numerically identified conformal multiplet (i.e. primary and its descendants) of the lowest $\mathrm{SO}(5)$ vector $\phi$, the lowest $\mathrm{SO}(5)$ traceless tensor $T$ and the symmetry current $J^\mu$ by matching the quantum numbers. For each system size $N_\mathrm{orb}$, the data is measured at a size-dependent parameter value $V/U$ around $0.9$ such that $\Delta_\mathscr{T}=3$ holds exactly, which is another requirement of conformal symmetry. We are able to find all their descendants up to $\ell\le 3$ and $\Delta\le 5$ with none missing. In general, the finite size effect is larger for a larger $\ell$, a similar behavior has also been observed for the 3D Ising CFT on the fuzzy sphere~\cite{Zhu2023Uncovering}. The measured scaling dimensions (symbols) and the corresponding anticipated values (grey lines) exhibit good agreement. More conformal multiplets like these are summarized in Table~\ref{tbl:full_spec} in the appendix, which contains 23 primaries and 76 conformal descendants. These results convincingly demonstrate the emergent conformal symmetry.

\begin{figure}[b!]
    \centering
    \includegraphics[width=\linewidth]{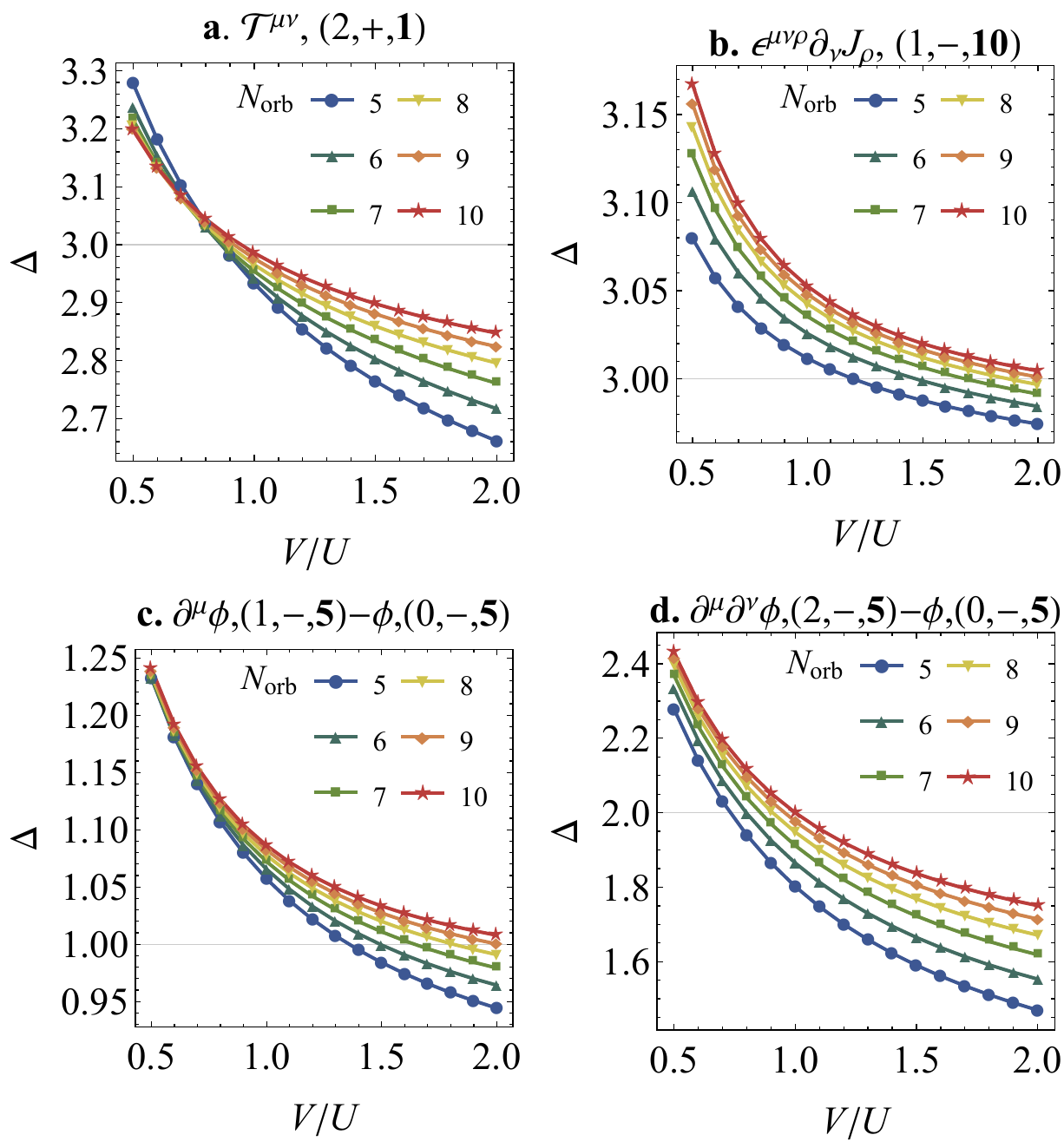}
    \caption{The scaling dimension of (a) stress tensor $\mathscr{T}^{\mu\nu}$, (b) the descendant $\epsilon^{\mu\nu\rho}\partial_\nu J_\rho$ of the symmetry current, (c,d) the diffence between the lowest $\mathrm{SO}(5)$ vector $\phi$ and its descendants (c) $\partial^\mu\phi$ and (d) $\partial^\mu\partial^\nu\phi$ at different system size $N_\mathrm{orb}$ calibrated by the scaling dimension of the symmetry current $\Delta_J=2$. The quantum numbers $(\ell,\mathscr{P},\mathrm{rep.})$ are given in the bracket. The grey gridline indicates the values imposed by the conformal symmetry.}
    \label{fig:4}
\end{figure}

We remark that certain intervals do not have the expected trend when increasing the system size (e.g., $\Delta_{\partial^\mu\phi}-\Delta_\phi$ scales to approximately $1.1$). This may come from either an insufficiently large system size or the lack of exact conformal symmetry due to the pseudo-criticality (see discussion below). It is also worth noting that if we move away from this parameter point, approximate conformal symmetry still holds in a vast region. We examine scaling dimension of the stress tensor $\mathscr{T}^{\mu\nu}$ and the descendants $\partial^\mu\phi$, $\partial^\mu\partial^\nu\phi$ and $\epsilon^{\mu\nu\rho}\partial_\nu J_\rho$ and compare them with the anticipated values (Fig.~\ref{fig:4}). For $\mathscr{T}^{\mu\nu}$, the agreement holds with a maximal discrepancy of $5\%$ for a vast region $0.7<V/U<1.5$, and the error decreases with increasing $N_\mathrm{orb}$; for $\epsilon^{\mu\nu\rho}\partial_\nu J_\rho$, the agreement holds with a maximal discrepancy of $3\%$.

\begin{figure}[t!]
    \centering
    \includegraphics[width=\linewidth]{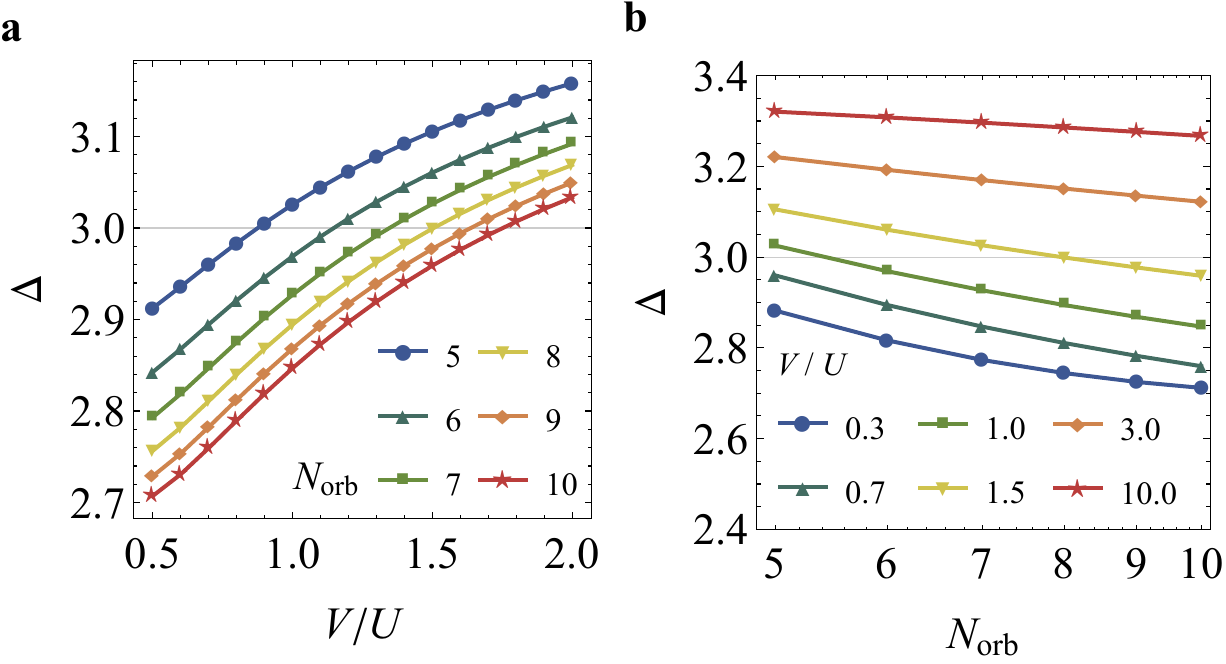}
    \caption{The scaling dimension of the lowest scalar $S$ (a) as a function of $V/U$ for different $N_\mathrm{orb}$ and (b) as a function of $N_\mathrm{orb}$ for different $V/U$. The horizontal axis of (b) is taken in logarithmic scale to compare with Fig.~\ref{fig:1_1}c. The grey gridline indicates $\Delta=3$ that separates relevance and irrelevance.}
    \label{fig:4_2}
\end{figure}

Our observation provides strong support for an approximate conformal symmetry in a vast region $0.7<V/U<1.5$. The next question we want to answer is whether this CFT signature corresponds to the fixed point $B$ (a phase transition), $C$ (a genuine $\mathrm{SO}(5)$ CFT) or $Z$ (pseudo-criticality) in Fig.~\ref{fig:1}b. The vast region of approximate conformal symmetry rules out the possibility of an unstable fixed point $B$ that needs fine-tuning. The distinction between a genuine $\mathrm{SO}(5)$ CFT and pseudo-criticality can be further diagnosed by the lowest singlet $S$. In the case of real fixed points, $\Delta_S$ will increase towards irrelevance with $\Delta_S > 3$, while for pseudo-criticality, $\Delta_S$ will decrease from irrelevant ($\Delta_S > 3$) towards relevant ($\Delta_S < 3$) along the flow.
Our results indeed observe such flow (Fig.~\ref{fig:4_2} and Table~\ref{tbl:2}).
In particular, for $1.0\lesssim V/U\lesssim 1.5$, $\Delta_S$ flows from slightly irrelevant ($\Delta_S\gtrsim3$) to slightly relevant ($\Delta_S\lesssim 3$) as the system size increases, supporting the scenario that DQCP corresponds to not a real CFT, but to a pseudo-critical region that locates near complex CFT fixed points and exhibits approximate conformal symmetry. 

\begin{table}[t!]
    \centering
    \setlength{\tabcolsep}{6pt}
    \caption{The scaling dimensions $\Delta_S$ of the lowest singlet $S$ at different $V/U$ and system size $N_\mathrm{orb}$ calibrated by the scaling dimension of the symmetry current $\Delta_J=2$.}
    \begin{tabular}{r|cccccc}
        \hline\hline
        \multirow{2}{*}{$V/U$}&\multicolumn{6}{c}{$N_\mathrm{orb}$}\\
        &10&9&8&7&6&5\\
        \hline
        $0.3$ & $2.712$&$2.725$&$2.744$&$2.773$&$2.817$&$2.882$\\
        $0.7$ & $2.760$&$2.782$&$2.811$&$2.847$&$2.895$&$2.960$\\
        $0.9$ & $2.819$&$2.841$&$2.868$&$2.902$&$2.946$&$3.005$\\
        $1.0$ & $2.847$&$2.868$&$2.894$&$2.927$&$2.969$&$3.026$\\
        $1.5$ & $2.959$&$2.977$&$2.999$&$3.026$&$3.061$&$3.106$\\
        $3.0$ & $3.122$&$3.136$&$3.151$&$3.170$&$3.193$&$3.221$\\
        $10.0$& $3.267$&$3.276$&$3.286$&$3.297$&$3.308$&$3.321$\\
        \hline\hline
    \end{tabular}
    \label{tbl:1}
\end{table}

\subsection{Operator spectrum}
\label{sec:3d}

\begin{figure}[b!]
    \centering
    \includegraphics[width=\linewidth]{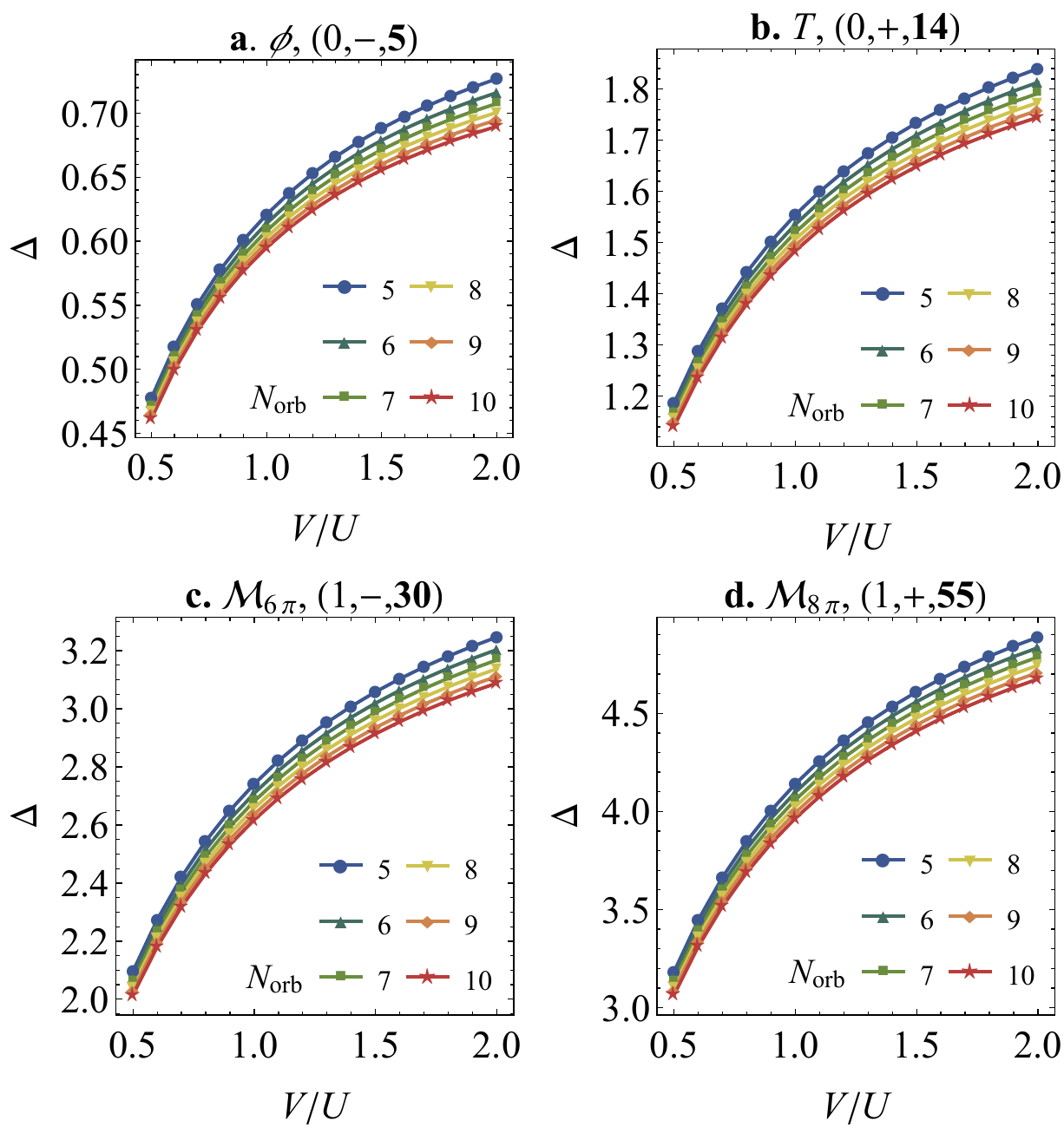}
    \caption{The scaling dimension of (a) the lowest $\mathrm{SO}(5)$ vector $\phi$, (b) the lowest rank-2 $\mathrm{SO}(5)$ symmetric traceless tensor $T$, (c) the $6\pi$-monopole $\mathscr M_{6\pi}$ and (d) the $8\pi$-monopole $\mathscr M_{8\pi}$ as a function of $V/U$ at different system size $N_\mathrm{orb}$ calibrated by the scaling dimension of the symmetry current $\Delta_J=2$. The quantum numbers $(\ell,\mathscr{P},\mathrm{rep.})$ are given in the bracket.}
    \label{fig:5}
\end{figure}

Having presented the evidence that DQCP is likely pseudo-critical, we now turn to its (pseudo-)critical properties, i.e., scaling dimensions of primary operators. Since there are no true CFT fixed points in the real axis, the operator spectrum presented below should be viewed as the real part of the complex scaling dimensions of the true complex CFTs. We also note that the scaling dimensions of the operators change with the parameter $V/U$ (Fig.~\ref{fig:5}), this may be the result of the walking behavior in the vicinity of the complex fixed point. The parameter dependence of scaling dimensions of various primaries appears to follow the same pattern, suggesting a potential universal behavior that can be understood through RG analysis.

As we target the pseudo-critical region in the RG flow diagram, to minimize the finite-size effect, for each system size $N_\mathrm{orb}$, we conduct the calculation at a $V/U$ value where $\Delta_\mathscr{T}=3$ holds exactly. We analyze the operators through the following process: (1) We pick out the lowest state in each representation and identify it as a primary; (2) we identify its descendants by matching the quantum numbers $(\ell,\mathscr{P},\mathrm{rep.};\Delta)$; (3) we remove the identified conformal multiplet from the spectrum and repeat the process. The lowest-lying primaries are listed in Table~\ref{tbl:2}. We complement this table with two other operators $\mathscr M_{8\pi}$ and $S^-$ which will be explained in the following. The full spectrum can be found in Appendix~\ref{sec:app_spec}.

Besides $J^\mu, \mathscr{T}^{\mu\nu}$ and $S$ that we have introduced before, there are several other primary operators worth noting:

\begin{figure}[b!]
    \centering
    \includegraphics[width=\linewidth]{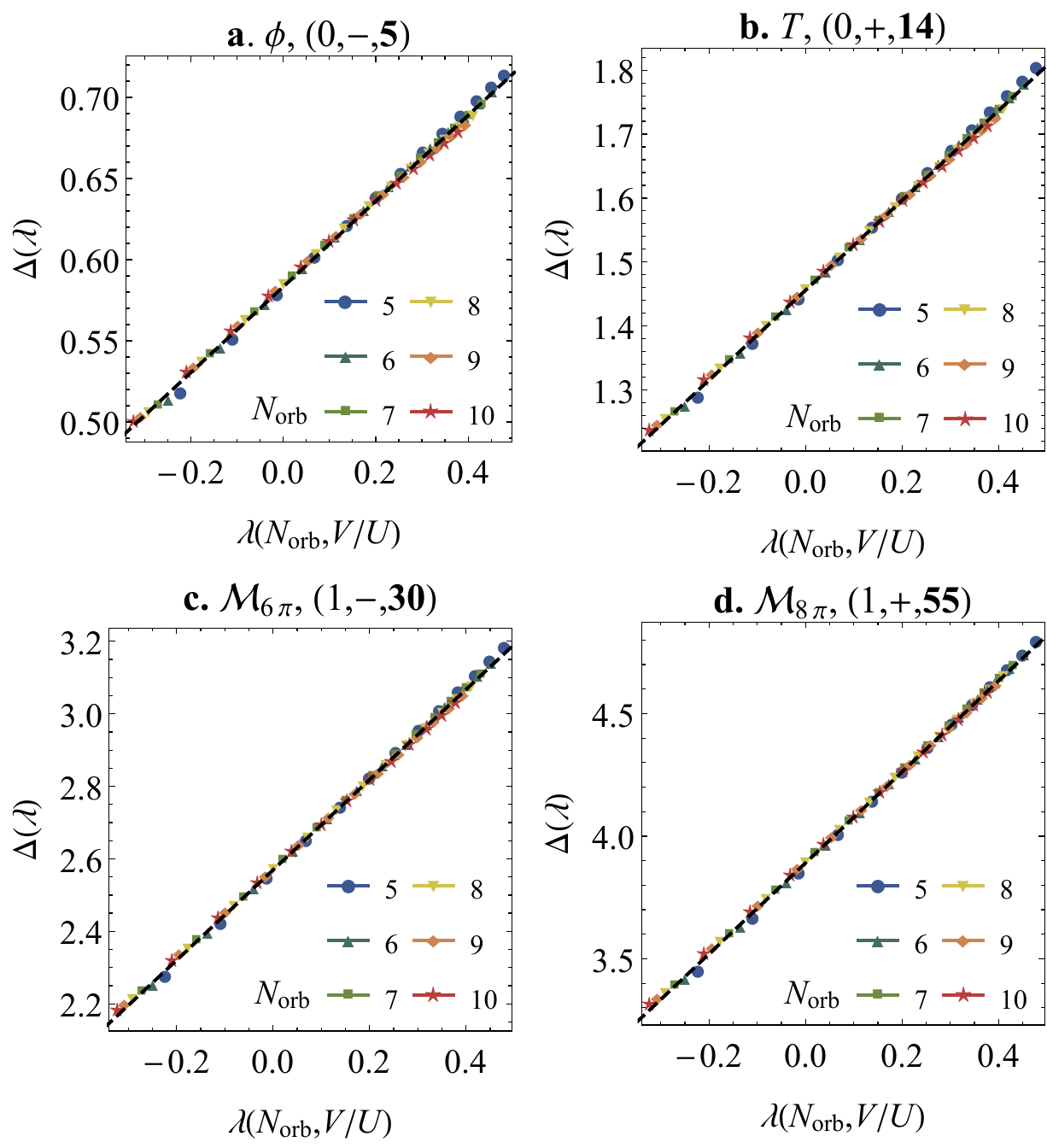}
    \caption{The data collapse for the scaling dimensions as a universal function of the running coupling parameter $\lambda(N_\mathrm{orb},V/U)$ Eq. \eqref{eq:running_coupling} for different system sizes. The data are the same as Fig.~\ref{fig:5}. The black line on each figure denotes the fitted result. Note that the $x$-axis $\lambda(N_\mathrm{orb},V/U)$ could differ up to an arbitrary linear transformation, and the zero point of $\lambda$ is taken such that $\Delta_{\mathscr{T}^{\mu\nu}}=3$ holds exact at $N_\mathrm{orb}=10$.}
    \label{fig:5_1}
\end{figure}

\begin{table}[t!]
    \centering
    \caption{The scaling dimension and quantum numbers for the lowest lying primary operators obtained from state-operator correspondence at different system sizes $N_\mathrm{orb}$. These numbers still violate bootstrap bound. Parameter $V/U$ is taken such that $\Delta_\mathscr{T}=3$ exactly.}
    \setlength{\tabcolsep}{3.9pt}
    \begin{tabular}{c|ccc|ccccc}
        \hline\hline
        &\multicolumn{3}{c|}{$N_\mathrm{orb}$}&10&9&8&7&6\\
        &\multicolumn{3}{c|}{$V/U$}&$0.9437$&$0.9150$&$0.8904$&$0.8717$&$0.8617$\\
        \hline
        Op.&$\ell$ & $\mathscr{P}$ & Rep. &&& $\Delta$ && \\\hline
        $\mathbb{I}$            & $0$ & $+$ & $\mathbf{1 }$ & $0.000$ & $0.000$ & $0.000$ & $0.000$ & $0.000$ \\
        $\phi$                  & $0$ & $-$ & $\mathbf{5 }$ & $0.585$ & $0.584$ & $0.583$ & $0.583$ & $0.586$ \\
        $T$                     & $0$ & $+$ & $\mathbf{14}$ & $1.458$ & $1.454$ & $1.452$ & $1.455$ & $1.463$ \\
        $J^\mu$                 & $1$ & $+$ & $\mathbf{10}$ & $2.000$ & $2.000$ & $2.000$ & $2.000$ & $2.000$ \\
        $\mathscr M_{6\pi}$     & $0$ & $-$ & $\mathbf{30}$ & $2.571$ & $2.565$ & $2.562$ & $2.567$ & $2.582$ \\
        $S$                     & $0$ & $+$ & $\mathbf{1 }$ & $2.831$ & $2.845$ & $2.865$ & $2.894$ & $2.937$ \\
        $\mathscr{T}^{\mu\nu}$  & $2$ & $+$ & $\mathbf{1 }$ & $3.000$ & $3.000$ & $3.000$ & $3.000$ & $3.000$ \\
                                & $1$ & $-$ & $\mathbf{35}$ & $3.031$ & $3.028$ & $3.028$ & $3.030$ & $3.037$ \\
                                & $1$ & $+$ & $\mathbf{10}$ & $3.164$ & $3.167$ & $3.171$ & $3.176$ & $3.183$ \\
                                & $2$ & $+$ & $\mathbf{14}$ & $3.333$ & $3.330$ & $3.325$ & $3.315$ & $3.283$ \\\hline
        $\mathscr M_{8\pi}$     & $0$ & $+$ & $\mathbf{55}$ & $3.895$ & $3.885$ & $3.881$ & $3.887$ & $3.908$ \\
        $S^-$                   & $0$ & $-$ & $\mathbf{1 }$ & $5.338$ & $5.354$ & $5.366$ & $5.373$ & $5.372$ \\
        \hline\hline
    \end{tabular}
    \label{tbl:2}
\end{table}

\begin{enumerate}[label=(\arabic*)]
    \item The lowest $\ell=0$ parity-odd $\mathrm{SO}(5)$ vector $\phi$ corresponds to the order parameter. Its scaling dimension is related to the anomalous dimension $\eta=(\Delta_\phi-1/2)/2$. Our $N_\mathrm{orb}=10$ data implies $\eta=0.168$.
    \item The lowest $\ell=0$ parity-even symmetric rank-2 tensor $T$ corresponds to the relevant perturbation that controls the original N\'eel-VBS transition. Its scaling dimension is related to the exponent $\nu=1/(3-\Delta_T)$. Our $N_\mathrm{orb}=10$ data implies $\nu=0.647$.
    \item The lowest $\ell=0$ parity-odd operator in $\ydiagram{3}$ representation corresponds to the $6\pi$-monopole $\mathscr M_{6\pi}$ in the CP$^1$ description. This operator is forbidden in lattices with $C_4$ rotation symmetry but allowed for $C_3$ symmetry. Although the exact value of its scaling dimension flows (Fig. \ref{fig:5}c), our calculation finds it relevant in all cases, which is likely to imply that the DQCP is not possible on honeycomb lattice as $\mathscr M_{6\pi}$ will drive it away~\cite{Sumiran2013Neel,Sato2017Dirac}.
    \item The lowest $\ell=0$ parity-even operator in $\ydiagram{4}$ representation corresponds to the $8\pi$-monopole $\mathscr M_{8\pi}$ in the CP$^1$ description. This operator is related to the dangerous irrelevant perturbation in the N\'eel-VBS DQCP~\cite{Senthil2004Deconfined,Senthil2004Quantum}. Although the exact value of its scaling dimension flows (Fig. \ref{fig:5}d), our calculation confirms its irrelevance.
    \item The lowest $\ell=0$ parity-odd singlet $S^-$ corresponds to the fermion bilinear in the QCD$_3$ description. This operator has engineering dimension $2$, but receives a huge correction up to $\Delta_{S^-}\approx 5.37$~\footnote{In the CP$^1$ description, this operator has a large engineer dimension.}. If this operator were relevant, it would drive the DQCP towards a chiral spin liquid, potentially playing a role in interesting phenomena observed in real materials~\cite{samajdar2019enhanced}. Our calculation finds it highly irrelevant and therefore negates this scenario.
\end{enumerate}

\begin{figure*}[t!]
    \centering
    \includegraphics[width=\linewidth]{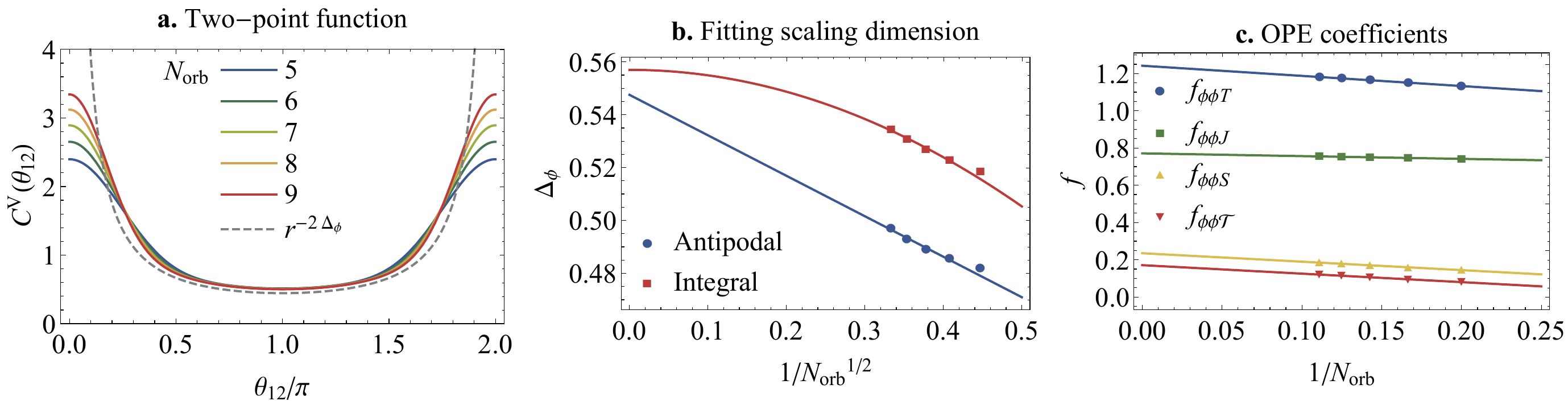}
    \caption{(a) The dimensionless two-point correlation function $C^\mathrm{V}(\theta_{12})$ defined in Eq.~(\ref{eq:3d_2pt}) at different system size $N_\mathrm{orb}$ and the theoretical expectation in the thermodynamic limit $C(\theta_{12})=(2\sin\frac{\theta_{12}}{2})^{-2\Delta_\phi}$; (b) the scaling dimension $\Delta_\phi$ extrapolated from its value at antipodal points Eq.~(\ref{eq:3d_antip}) and spatial integral Eq.~(\ref{eq:3d_int}); (c) The finite-size scaling of OPE coefficients $f_{\phi\phi T}$, $f_{\phi\phi J}$, $f_{\phi\phi S}$ and $f_{\phi\phi\mathscr{T}}$.}
    \label{fig:6}
\end{figure*}

It is worth noting that the scaling dimensions of various monopoles in the CP$^1$ description have been calculated by introducing such defects into 3D critical dimer model in Monte Carlo simulation~\cite{Sreejith2015Monopoles,Sreejith2014Monopoles}, including $\Delta_\phi=0.579(8)$, $\Delta_T=1.52(7)$ and $\Delta_{\mathscr M_{6\pi}}=2.80(3)$, which are very close to our results and also confirm the relevance of $6\pi$-monopole. We can also compare the critical exponents with the Monte Carlo results in various transitions, including 3D loop model $\eta_{\textrm{loop, VBS}}=0.25(3)$, $\eta_{\textrm{loop, N\'eel}}=0.259(3)$, $\nu_{\textrm{loop, VBS}}=0.503(9)$ and $\nu_{\textrm{loop, N\'eel}}=0.477(4)$~\cite{Nahum2015Deconfined}, $J$-$Q$ model $\eta_{\textrm{$J$-$Q$}}=0.35$, $\nu_{\textrm{$J$-$Q$}}=0.455(2)$~\cite{Sandvik2010Continuous,Shao2016Quantum,Sandvik2020Consistent} and transition between quantum spin Hall insulators and $s$-wave superconductor (QSH-SC)~\cite{Grover2008Topological} $\nu_\textrm{SC}=0.56(6)$, $\nu_\textrm{QSH}=0.6(1)$, $\eta_\textrm{SC}=0.22(6)$ and $\eta_\mathrm{QSH}=0.21(5)$~\cite{Liu2019Superconductivity}. We think that the discrepancy between these results is a consequence of critical exponents drift of walking behavior of pseudo-criticality. Phenomenologically, the exponents computed using the RG analysis (or scaling assumptions) of true criticality will be drifting with the system size and interaction strengths, as shown in Fig.~\ref{fig:5}. Similar drift with system size has also been observed in $\nu$ in loop models~\cite{Nahum2015Deconfined} and both $\eta$ and $\nu$ in the QSH-SC transition~\cite{Liu2019Superconductivity}. For example, in the loop model computation~\cite{Nahum2015Deconfined}, $\nu$ is drifting from $\nu\approx 0.65$ at $L\sim 50$ to $\nu\approx 0.50$ at $L\sim 500$. In the next section we will show that our data of scaling dimensions drift can be understood using the conformal perturbation of the pseudo-criticality discussed in Sec.~\ref{sec:2b_1}.

\subsection{A pseudo-critical data collapse for scaling dimension drifts}
\label{sec:3d_1}

To verify that the scaling dimensions drift (Fig.~\ref{fig:5}) is the consequence of the walking behavior in the vicinity of the complex fixed point, we perform a data collapse based on the theory of pseudo-criticality. We try to fit the parameter and size dependence with the prediction of conformal perturbation Eqs.~\eqref{eq:running_coupling} and \eqref{eq:perturb_final}. For the dependence of $\lambda_0$ on $V/U$, we assume that both $H_\mathrm{CFT}$ and $S$ can be expressed as a linear combination of two terms in the Hamilonian $H_\mathrm{CFT}=\int\mathrm{d}^2\vec{r}(U_0n^2-V_0\Delta^\dagger\Delta)$ and $S=U_1n^2-V_1\Delta^\dagger\Delta$. The $\lambda_0(V/U)$ then take the ansatz $\lambda_0(V/U)=C_1(1+C_2V/U)/(1+C_3V/U)$. To facilitate fitting we linearise Eq.~\eqref{eq:running_coupling} into $\lambda(R,\lambda_0)=\lambda_0-\alpha y^2\log(R/R_0)$. Under these assumptions, the fitting ansatz can be organized into
\begin{align}
    \Delta_\Phi(N_\mathrm{orb},V/U)&=\Delta_{\Phi,0}+f_{\Phi\Phi S}\lambda(N_\mathrm{orb},V/U)\\
    \lambda(N_\mathrm{orb},V/U)&=C_1\frac{1+C_2V/U}{1+C_3V/U}-\frac{1}{2}\alpha y^2\log N_\mathrm{orb}
\end{align} 
with the constants $\Delta_{\Phi,0},C_1f_{\Phi\Phi S},\alpha y^2/C_1$ and $C_{2,3}$ to be fitted and the linear size is taken as $R=\sqrt{N_\mathrm{orb}}$. We find that the scaling dimensions plotted as a function of the running coupling $\lambda(N_\mathrm{orb},V/U)$ for different system sizes collapse well (Fig.~\ref{fig:5_1}), which is further evidence for the scenario of pseudo-criticality. The goodness of the fittings can be evaluated by the average relative residual $\delta_\Phi=\langle(\Delta^\mathrm{ansatz}_\Phi(N_\mathrm{orb},V/U)/\Delta^\mathrm{numerics}_\Phi(N_\mathrm{orb},V/U)-1)^2\rangle^{1/2}$, which are around $\delta\approx3.5\times10^{-3}$ for all the quantities in Fig.~\ref{fig:5_1}. In addition, if we make a further assumption that $\Delta_\mathscr{T}=3$ holds exactly at the point $\lambda=0$, we can fit the scaling dimensions $\Delta_{\Phi,0}$: $\Delta_\phi=0.583(7),\Delta_T=1.456(12),\Delta_{\mathscr{M}_{6\pi}}=2.569(19),\Delta_{\mathscr{M}_{8\pi}}=3.89(3),\Delta_S=2.90(6),\Delta_{\mathscr{T}^{\mu\nu}}=3.00(6)$. 

We also need to stress that these values are calculated in the walking region off the critical point. To produce reliable scaling dimensions, one should consider a non-Hermitian Hamiltonian and tune parameters to the complex fixed points. We also need to stress that the analysis of conformal perturbation above is very preliminary. The subleading contributions from the irrelevant operators have not been eliminated. The undetermined parameters in the fitting may also be determined by measurements at the complex fixed point. 

\subsection{Correlation functions and OPE coefficients}
\label{sec:3e}

Having studied the spectrum of the system and the corresponding CFT, we now turn to the operators and their correlation functions. The simplest particle-hole symmetric operator is the density operator~\cite{Hu2023Operator}
\begin{equation}
    \hat{n}^\mathbf{M}(\theta,\varphi)=\hat{\mathbf{\Psi}}^\dagger(\theta,\varphi)\mathbf{M}\hat{\mathbf{\Psi}}(\theta,\varphi),
\end{equation}
where $\mathbf{M}$ is an Hermitian matrix insertion. Any gapless density operator in the microscopic model can be expressed as a linear combination of CFT operators including primaries and descendants that have the same parity and $\mathrm{SO}(5)$ quantum number as $\hat{n}^\mathbf{M}$
\begin{equation}
    \hat{n}^\mathbf{M}(\theta,\varphi;\tau=0)=\sum_\alpha c_\alpha\hat{\Phi}_\alpha.
\end{equation}
Using this decomposition, we may consider the one-point functions
\begin{align}
    \langle\Phi_\alpha|\hat{n}^\mathbf{M}(\theta,\varphi;\tau=0)|0\rangle&=\sum_{\beta\in[\alpha]}c_\beta R^{-\Delta_\beta} h_\beta(\theta,\varphi),\nonumber\\
    \langle\Phi_\alpha|\hat{n}^\mathbf{M}(\theta,\varphi;\tau=0)|\Phi_\gamma\rangle&=\sum_{\beta}f_{\alpha\beta\gamma}c_\beta R^{-\Delta_\beta}\tilde{h}_{\alpha\beta\gamma}(\theta,\varphi),
\end{align}
where $|\Phi_\alpha\rangle$ denotes the state corresponding to the CFT operator $\hat{\Phi}_\alpha$, $[\alpha]$ denotes conformal multiplet of $\alpha$, $h_\beta(\theta,\varphi)$ and $\tilde{h}_{\alpha\beta\gamma}(\theta,\varphi)$ are universal functions fixed by conformal symmetry, and $f_{\alpha\beta\gamma}$ is the OPE coefficient. Specifically, we may consider the density operator in the vector representation of $\mathrm{SO}(5)$ (i.e., the $\mathrm{Sp}(2)$ antisymmetric rank-$2$ traceless tensor representation) by inserting the $\gamma$-matrices
\begin{equation}
    \hat{n}^{\mathrm{V},i}(\theta,\varphi)=\hat{\mathbf{\Psi}}^\dagger(\theta,\varphi)\mathbf{\gamma}^i\hat{\mathbf{\Psi}}(\theta,\varphi).
\end{equation}
In the perspective of CFT, this operator receives its lowest contribution from the $\hat{\Phi}$ and its descendants $\hat{n}^\mathrm{V}=c_\phi\hat{\phi}+c_{\partial_\mu\phi}\partial_\mu\hat{\phi}+c_{\partial^2\phi}\partial^2\hat{\phi}+\dots$

The normalized two-point function of $\hat{n}^\mathrm{V}$ therefore receives its leading order contribution from the 2-pt function of $\hat{\phi}$
\begin{align}
    C^\mathrm{V}(\theta_1,\varphi_1;\theta_2,\varphi_2)&=C^\mathrm{V}(\theta_{12})\nonumber\\
    &=\frac{\langle 0|\hat{n}^\mathrm{V}(\theta_1,\varphi_1)\hat{n}^\mathrm{V}(\theta_2,\varphi_2)|0\rangle}{|\langle 0|\int\frac{\sin\theta\,\mathrm{d}\theta\,\mathrm{d}\varphi}{4\pi}\hat{n}^\mathrm{V}(\theta,\varphi)|\phi\rangle|^2}\nonumber\\
    &=\langle\hat{\phi}(\vec e_1)\hat{\phi}(\vec e_2)\rangle_\mathrm{flat}+\mathscr{O}(R^{-1})\nonumber\\
    &=(2\sin{\textstyle\frac{\theta_{12}}{2}})^{-2\Delta_\phi}+\mathscr{O}(R^{-1}),
    \label{eq:3d_2pt}
\end{align}
where $\theta_{12}$ is the angular distance and $\vec e_{1,2}$ are the unit vector of the two points, the subleading correction $\mathscr{O}(R^{-1})$ comes from the contribution of the descendant $\partial^\mu\hat{\phi}$ on the nominator. Numerically, we find the finite-size result approaches theoretical expectation as $N_\mathrm{orb}$ increases (Fig.~\ref{fig:6}a). At large distance near $\theta_{12}=\pi$, the finite-size result has little discrepancy with the theoretical expectation, while the divergence at small distance is not captured by the finite-size numerical result. From the 2-pt function, we can also extract the scaling dimension of $\phi$ by taking its value at antipodal points
\begin{equation}
    C^\mathrm V(\pi)=2^{-2\Delta_\phi}+\mathscr{O}(R^{-1})
    \label{eq:3d_antip}
\end{equation}
and its spatial integral
\begin{equation}
    \int\frac{\sin\theta\,\mathrm{d}\theta}{2}C^\mathrm{V}(\theta)=\frac{2^{-2\Delta_\phi}}{\Delta-1}+\mathscr{O}(R^{-2}),
    \label{eq:3d_int}
\end{equation}
where for the latter the subleading contribution comes from $\partial^2\hat{\phi}$ instead. After a finite-size scaling, we extrapolate that $\Delta_\phi^{(\textrm{antipodal})}=0.56(2)$ and $\Delta_\phi^{(\textrm{integral})}=0.55(5)$ (Fig.~\ref{fig:6}b), which are $6\%$ and $5\%$ different from the result from the state-operator correspondence $0.584$.

\begin{table}[b!]
    \centering
    \setlength{\tabcolsep}{10pt}
    \caption{The extrapolated OPE coefficients. The error bar is extracted from linear extrapolation.}
    \label{tbl:3}
    \begin{tabular}{cccc}
        \hline\hline
        $f_{\phi\phi T}$&$f_{\phi\phi J}$&$f_{\phi\phi S}$&$f_{\phi\phi\mathscr{T}}$\\
        \hline
        $1.242(7)$&$0.771(3)$&$0.235(8)$&$0.121(8)$\\
        \hline\hline
    \end{tabular}
\end{table}

On the other hand, we can extract the OPE coefficients by taking the inner product of $\hat{n}^\mathrm{V}(\theta,\varphi)$ with CFT states and integrate out the angular dependence~\cite{Hu2023Operator}. For details see Appendix~\ref{sec:app_ope}. As an example,
\begin{multline}
   f_{\phi\phi J}=\sqrt\frac{3}{2}\frac{\int\frac{\sin\theta\,\mathrm{d}\theta\,\mathrm{d}\varphi}{4\pi}\bar{Y}_{10}(\theta,\varphi)\langle\phi|n^\mathrm{V}(\theta,\varphi)|J_{m=0}\rangle}{\int\frac{\sin\theta\,\mathrm{d}\theta\,\mathrm{d}\varphi}{4\pi}\bar{Y}_{00}(\theta,\varphi)\langle\phi|n^\mathrm{V}(\theta,\varphi)|0\rangle}\\+\mathscr{O}(R^{-2}).
\end{multline}
where the subleading contribution comes from $\partial^2\hat{\phi}$. Similarly, we calculate several OPE coefficients. After a finite-size scaling (Fig.~\ref{fig:6}c), the extrapolated values are listed in Table~\ref{tbl:3}. Here we use a convention that the two-point correlator of $J^\mu$ or $\mathscr{T}^{\mu\nu}$ normalizes to $1$, so OPE coefficients $f_{\phi\phi J}$ and $f_{\phi\phi\mathscr{T}}$ can  give central charges. For example, the stress tensor central charge $C_{\mathscr{T}}=(\frac{3 \Delta_\phi}{4f_{\phi\phi\mathscr{T}}})^2\approx 6.561 = 0.8748 \cdot (5 C_{\mathscr{T}}^{\textrm{free}}) $, where $C_{\mathscr{T}}^{\textrm{free}}=1.5$ is the central charge of a free real scalar~\cite{Poland2019Conformal}.

\section{Summary and discussions}
\label{sec:4}

In this paper, we utilize the fuzzy sphere regularization as a microscope to investigate the $\mathrm{SO}(5)$ NL$\sigma$M with a level-1 WZW term, which serves as one of the dual descriptions of the $\mathrm{SO}(5)$ DQCP. We present compelling evidence supporting the presence of an approximate conformal symmetry in the model. Specifically, in the excitation spectrum, we have identified many characteristic features of emergent conformal symmetry, including the conserved $\mathrm{SO}(5)$ symmetry current, the stress tensor, and observed integer-spaced levels between primary operators and their descendants. Furthermore, we find that the renormalization group (RG) flow supports the scenario of pseudo-criticality. In particular, we observe the lowest symmetry singlet flowing from being slightly irrelevant to slightly relevant, which is a characteristic feature of pseudo-criticality. The scaling dimensions of other primaries also exhibit size and parameter dependences that are quantitatively consistent with the prediction of pseudo-criticality. Additionally, we identify various primary operators, including a relevant $6\pi$-monopole (in the context of the CP$^1$ model), an irrelevant $8\pi$-monopole, and a highly irrelevant parity odd $\mathrm{SO}(5)$ singlet. These findings hold important physical implications. Furthermore, we have computed several OPE coefficients, including the central charge of the stress tensor.

It is worth noting that our work is the first time observing that the $\mathrm{SO}(5)$ DQCP has approximate conformal symmetry. This essentially rules out several proposals explaining the abnormal scaling behavior in the N\'eel-VBS transition, e.g., the N\'eel-VBS transition is described by a continuous transition without conformal symmetry. We also need to add that, in principle, the subleading contributions from multiple irrelevant operators in the vicinity of a real fixed point could cause a similar flow behavior of the singlet dimension decreasing towards relevance. Although we have not found any indication of it, our analysis currently cannot rule out this small possibility. To rule out this scenario, a more sophisticated analysis with conformal perturbation~\cite{Lao20233Icosahedron} taking into account the irrelevant perturbations of the Hamiltonian can be carried out in the future. Besides, other techniques such as density matrix renormalization group (DMRG)~\cite{Hu2023Operator} and quantum Monte Carlo (QMC)~\cite{Hofmann2023Quantum} can also be conveniently applied to fuzzy sphere and greatly expand the system sizes. With the help of these methods, our numerical findings could be further strengthened in the future.

So far, our exploration has been limited to the pseudo-critical phenomenon, which serves as a shadow of the complex fixed point in the complex plane. It is highly intriguing to directly investigate the physics of the complex fixed point itself. Establishing its existence and comprehending its nature would not only conclusively settle the two-decade-long debate surrounding the DQCP but also provide fresh insights into the landscape of fixed points and CFTs, which hold fundamental significance. It is conjectured that the complex fixed point is a relatively common feature in many models and theories, representing one of the few, if not the only, known mechanisms for interaction-driven first-order phase transitions. However, apart from a few examples in 2D (i.e., $(1+1)$D), no example in 3D or higher dimensions has been firmly established thus far. The study of the complex fixed point necessitates the examination of a non-Hermitian Hamiltonian, a task made feasible through the fuzzy sphere technique.

We also emphasize that the observation of pseudo-critical behavior in the DQCP should not diminish its significance. Pseudo-criticality closely resembles true criticality over a wide range of length scales (e.g., system size) or energy scales (e.g., temperature). For instance, any experimental realization of a quantum phase transition is necessarily conducted at a finite temperature, so for a pseudo-critical system one would also observe universal phenomena governed by the complex fixed point. Therefore, employing the fuzzy sphere technique to uncover the CFT perspective of the DQCP at finite temperature, an aspect inaccessible through traditional lattice model simulations, holds great intrigue.

In addition to observing pseudo-critical behavior, we have demonstrated the efficacy of the fuzzy sphere microscope by computing the scaling dimensions of many primary operators. These results play a vital role in enhancing our understanding of the DQCP in various systems, which were previously unattainable through earlier studies. For instance, our findings indicate that the $\mathrm{SO}(5)$ DQCP cannot be applied to the N\'eel-VBS transition on the honeycomb lattice. Consequently, it becomes imperative to employ the fuzzy sphere microscope in investigating other intricate CFTs. One primary target of interest is the $\mathrm{U}(1)$ Dirac spin liquid~\cite{Affleck1988,Wen1995,Hermele2005}, whose comprehension holds significant value for experimental studies involving real materials. Specifically, it is crucial to determine the (ir)relevance of specific operators, as this determines whether the $\mathrm{U}(1)$ Dirac spin liquid represents a stable phase of matter or a phase transition on triangular or kagome lattices and related materials~\cite{Jian2018,Song2018,Song2018a}.

Another exciting application of the fuzzy sphere microscope is to solve the conformal window problem of 3D critical gauge theories, a long-standing open problem that is interesting to both condensed matter and high-energy physics. Specifically, the $\mathrm{SO}(5)$ DQCP studied here is dual to $N=2$ Dirac fermions coupled to an $\mathrm{SU}(2)$ gauge field. We have also provided a simple model for its large-$N$ generalization, which corresponds to the QCD$_3$ theory with $2N$ Dirac fermions coupled to an $\mathrm{SU}(2)$ gauge field. By studying this model on the fuzzy sphere, one should be able to determine the precise region of the conformal window (i.e., $N>N_c$) for which the QCD$_3$ theory becomes conformal. The traditional methods, such as lattice model simulations~\cite{Karthik2018QCD3}, may not be able to complete this task due to the challenge of distinguishing a true critical (conformal) theory from a pseudo-critical theory. Moreover, generalizing this scheme to other critical gauge theories with different gauge groups should be feasible and interesting to explore in the future.

\begin{acknowledgments}
    We would like to thank Subir Sachdev, Chong Wang and Xue-Feng Zhang for illuminating discussions. Z.Z. acknowledges support from the Natural Sciences and Engineering Research Council of Canada (NSERC) through Discovery Grants. Research at Perimeter Institute is supported in part by the Government of Canada through the Department of Innovation, Science and Industry Canada and by the Province of Ontario through the Ministry of Colleges and Universities. LDH and WZ were supported by National Natural Science Foundation of China (No.~92165102, 11974288) and National key R\&D program (No. 2022YFA1402204).
\end{acknowledgments}

\appendix

\section{Sectionning the Hilbert space}
\label{sec:app_qn}

In the exact algorithm, we consider the following $\mathrm{U}(1)$ conserved quantities
\begin{equation}
    \hat{m}^z=\sum_mm\hat{\mathbf{c}}_m^\dagger\hat{\mathbf{c}}_m, \ \hat{\sigma}_1=\sum_m\hat{\mathbf{c}}_m^\dagger\boldsymbol{\sigma}_1\hat{\mathbf{c}}_m,\ \hat{\sigma}_2=\sum_m\hat{\mathbf{c}}_m^\dagger\boldsymbol{\sigma}_2\hat{\mathbf{c}}_m
\end{equation}
where
\begin{equation*}
    \boldsymbol{\sigma}_1=\begin{pmatrix}
        1&0&0&0\\0&0&0&0\\0&0&-1&0\\0&0&0&0
    \end{pmatrix}\textrm{ and }\boldsymbol{\sigma}_2=\begin{pmatrix}
        0&0&0&0\\0&1&0&0\\0&0&0&0\\0&0&0&-1
    \end{pmatrix}.
\end{equation*}
These quantities are conserved because of the $\mathrm{SO}(3)$ rotation symmetry and the $\mathrm{SO}(5)$ flavor symmetry. These quantities divide the many-body Hilbert space into sectors and block-diagonalize the Hamiltonian. The $(m^z,\sigma_1,\sigma_2)=0$ representation can be further divided four $\mathbb{Z}_2$ symmetries, viz. parity $\mathscr{P}$, $\pi$-rotation around $y$-axis $\mathscr{R}_y$ and permutation of flavour indices $\mathscr{X}_{1,2}$. 
\begin{align}
    \mathscr{P}&:\hat{\mathbf{c}}_m\to \mathbf{J}(\hat{\mathbf{c}}^\dagger_m)^\mathrm{T},i\to-i\nonumber\\
    \mathscr{R}_y&:\hat{\mathbf{c}}_m\to \hat{\mathbf{c}}_{-m}\nonumber\\
    \mathscr{X}_1&:\hat{\mathbf{c}}_m\to \mathbf{J}\hat{\mathbf{c}}_m\nonumber\\
    \mathscr{X}_2&:\hat{c}_{m,1}\leftrightarrow\hat{c}_{m,2},\hat{c}_{m,3}\leftrightarrow\hat{c}_{m,4}
\end{align}
The eigenstates we obtain are also simultaneously eigenstates of these quantities. Using the branching rule $\mathfrak{so}(5)\supset\mathfrak{su}(2)\oplus\mathfrak{su}(2)$ ($\hat{\sigma}_{1,2}$ corresponds to the Cartan subalgebra of the $\mathfrak{su}(2)$s), we can list the degeneracy within each $(\sigma_1,\sigma_2)$ sector and each $(\sigma_1=0,\sigma_2=0,\mathscr{X}_1,\mathscr{X}_2)$ sector for each representation (Table~\ref{tbl:rep_deg}). By matching the degeneracy of the measured state, we can infer the representation of the corresponding operator.

\section{Full spectrum}
\label{sec:app_spec}

In Table~\ref{tbl:spec_scan}, we list the scaling dimensions of various operators at different $V/U$ and system size $N_\mathrm{orb}$. These results support the emergence of approximate conformal symmetry in a vast region $V/U\ge 0.7$.

We list the operator spectrum of operators with $\ell\leq 3$ and $\Delta<5.5$ measured at $N_\mathrm{orb}=8$ and $V/U=0.8904$ in Table~\ref{tbl:full_spec}, containing 2691 states corresponding to 137 operators, organized into different representations and conformal multiplets.

\section{OPE tensor structure}
\label{sec:app_ope}
The OPE coefficients are defined by the 2-pt and 3-pt functions, namely
\begin{align}
    \langle\phi_i(x_1)\phi_j(x_2)\rangle&=\delta_{ij}x_{12}^{-2\tau_\phi}\nonumber\\
    \langle T_{ij}(x_1)T_{kl}(x_2)\rangle&=T_{ij,kl}x_{12}^{-2\tau_T}\nonumber\\
    \langle\hat{J}_{ij}(x_1,z_1)\hat{J}_{kl}(x_2,z_2)\rangle&=A_{ij,kl}H(x_1,x_2,z_1,z_2)x_{12}^{-2\tau_J}\nonumber\\
    \langle S(x_1)S(x_2)\rangle&=x_{12}^{-2\tau_S}\nonumber\\
    \langle\hat{\mathscr{T}}(x_1,z_1)\hat{\mathscr{T}}(x_2,z_2)\rangle&=H(x_1,x_2,z_1,z_2)^2x_{12}^{-2\tau_\mathscr{T}}\\
    \langle\phi_i(x_1)\phi_j(x_2)T_{kl}(x_3)\rangle&=\frac{f_{\phi\phi T}T_{ij,kl}}{x_{12}^{2\tau_\phi-\tau_T}x_{23}^{\tau_T}x_{31}^{\tau_T}}\nonumber\\
    \langle\phi_i(x_1)\phi_j(x_2)\hat{J}_{kl}(x_3,z_3)\rangle&=\frac{f_{\phi\phi J}A_{ij,kl}V(x_1,x_2,x_3,z_3)}{x_{12}^{2\tau_\phi-\tau_J}x_{23}^{\tau_J}x_{31}^{\tau_J}}\nonumber\\
    \langle\phi_i(x_1)\phi_j(x_2)S(x_3)\rangle&=\frac{f_{\phi\phi S} \delta_{ij}} {x_{12}^{2\tau_\phi-\tau_S}x_{23}^{\tau_S}x_{31}^{\tau_S}}\nonumber\\
    \langle\phi_i(x_1)\phi_j(x_2)\hat{\mathscr{T}}(x_3,z_3)\rangle&=\frac{f_{\phi\phi\mathscr{T}}\delta_{ij}V(x_1,x_2,x_3,z_3)^2}{x_{12}^{2\tau_\phi-\tau_\mathscr{T}}x_{23}^{\tau_\mathscr{T}}x_{31}^{\tau_\mathscr{T}}},
\end{align}
where $\tau_\Phi=\Delta_\Phi+\ell_\Phi$; the indices $i,j,k,l$ are $\mathrm{SO}(5)$ indices and the tensor structures are given by
\begin{align}
    A_{ij,kl}&=\frac{1}{2}\delta_{ik}\delta_{jl}-\frac{1}{2}\delta_{il}\delta_{jk}\nonumber\\
    T_{ij,kl}&=\frac{1}{2}\delta_{ik}\delta_{jl}+\frac{1}{2}\delta_{il}\delta_{jk}-\frac{1}{5}\delta_{ij}\delta_{kl}.
\end{align}
and the conformal invariant tensors are~\cite{Costa2011Spinning}
\begin{align}
    H(x_1,x_2,z_1,z_2)&=\frac{1}{2}x_{12}^2(z_1\cdot z_2)-(z_1\cdot x_{12})(z_2\cdot x_{12})\nonumber\\
    V(x_1,x_2,x_3,z_3)&=\frac{1}{x_{12}^2}\left[x_{23}^2(z_3\cdot x_{13})-x_{13}^2(z_3\cdot x_{23})\right].
\end{align}
Here the Lorentz indices are treated in the index-free treatment where spinning operators are contracted with null auxiliary vector fields
\begin{equation}
    \hat{\Phi}_\ell(x,z)=\Phi^{\mu_1\dots\mu_\ell}(x)z_{\mu_1}\dots z_{\mu_\ell},\quad z^2=0
\end{equation}
and the Lorentz indices can be recovered by applying the stripping operator
\begin{equation}
    D_{z,\mu}=\frac{d-2}{2}\partial_{z_\mu}+z_\nu\partial_{z_\mu}\partial_{z_\nu}-\frac{1}{2}z_\mu\partial_{z_\nu}\partial_{z_\nu}.
\end{equation}

We now want to rewrite the correlators in terms of the inner products of states. Consider a spin-$\ell$ operator $\Phi_{\ell,R}$ in the $R$-representation of $\mathrm{SO}(5)$. Let
\begin{equation}
    |\Phi_\mathbf{n}^\mathbf{e}\rangle=\alpha(\Phi_\mathbf{n}^\mathbf{e})\lim_{x\to 0}n^{\mu_1\dots\mu_\ell}e_{ij\dots}\Phi_{\mu_1\dots\mu_\ell}^{ij\dots}(x)|0\rangle,
\end{equation}
where $n^{\mu_1\dots\mu_\ell}$ is the Lorentz polarization, and $e_{ij\dots}$ is the $\mathrm{SO}(5)$ polarization.
The coefficient $\alpha(\Phi_\mathbf{n}^\mathbf{e})$ is determined by the normalisation condition

\begin{widetext}
    \begin{equation}
        \langle\Phi_\mathbf{n}^\mathbf{e}|\Phi_\mathbf{n}^\mathbf{e}\rangle=|\alpha(\Phi_\mathbf{n}^\mathbf{e})|^2\lim_{x\to 0}(n^*)^{\mu'_1\dots\mu'_\ell}n^{\mu_1\dots\mu_\ell}e^*_{i'j'\dots}e_{ij\dots}\langle(\Phi_{\mu'_1\dots\mu'_\ell}^{i'j'\dots}(x))^\dagger\Phi_{\mu_1\dots\mu_\ell}^{ij\dots}(x)\rangle,
    \end{equation}
    where the conjugation in the radial quantization is taken as
    \begin{equation}
        (\Phi^\dagger)_{\mu_1\dots\mu_\ell}^{ij\dots}(x)=x^{2\Delta}I_{\mu_1}{}^{\nu_1}(x)\dots I_{\mu_\ell}{}^{\nu_\ell}(x)\Phi_{\nu_1\dots\nu_\ell}^{ij\dots}\left(\frac{x^\mu}{x^2}\right),
    \end{equation}
    where $I_\mu{}^\nu(x)=\delta_\mu^\nu-2x_\mu x^\nu/x^2$. Hence,
    \begin{equation}
        \alpha(\Phi_\mathbf{n}^\mathbf{e})=\left[\lim_{x\to\infty}x^{2\Delta_\Phi}(n^*)^{\mu'_1\dots\mu'_\ell}n^{\mu_1\dots\mu_\ell}e^*_{i'j'\dots}e_{ij\dots}I_{\mu'_1}{}^{\nu_1}(x)\dots I_{\mu'_\ell}{}^{\nu_\ell}(x)\langle\Phi_{\nu_1\dots\nu_\ell}^{i'j'\dots}(x)\Phi_{\mu_1\dots\mu_\ell}^{ij\dots}(0)\rangle\right]^{-1/2}.
    \end{equation}
\end{widetext}

\begin{table*}[p]
    \centering
    \setlength{\tabcolsep}{5pt}
    \caption{The Young diagrams and quadratic Casimir $C_2$ of different $\mathrm{Sp}(2)$ and $\mathrm{SO}(5)$ representations and the corresponding state degeneracies in different $(\sigma_1,\sigma_2)$ and $(\sigma_1=0,\sigma_2=0,\mathscr{X}_1,\mathscr{X}_2)$ sectors. Here we only listed the sectors with $0\le\sigma_2\le\sigma_1$. The sectors $(\pm\sigma_1,\pm\sigma_2)$ and $(\pm\sigma_2,\pm\sigma_1)$ should have the same degeneracy, e.g., the degeneracy listed for $(3,1)$ also applies to $(3,-1),(-3,\pm 1),(1,\pm 3)$ and $(-1,\pm 3)$.}
    \begin{tabular}{cllc|ccccccccc|cccc}
        \hline\hline
        rep.&\multicolumn{2}{c}{Young diagram}&$C_2$&\multicolumn{9}{c|}{Degeneracy in sector $(\sigma_1,\sigma_2)$}&\multicolumn{4}{c}{in sector $(0,0,\mathscr{X}_1,\mathscr{X}_2)$}\\
        &$\mathrm{Sp}(2)$&$\mathrm{SO}(5)$&&$(0,0)$&$(1,1)$&$(2,0)$&$(2,2)$&$(3,1)$&$(3,3)$&$(4,0)$&$(4,2)$&$(4,4)$&$(+,+)$&$(+,-)$&$(-,+)$&$(-,-)$\\
        \hline
        $\mathbf{1}  $&                &                &$0 $&1& & & & & & & & &1& & & \\
        $\mathbf{5}  $&$\ydiagram{1,1}$&$\ydiagram{1}$  &$2 $&1&1& & & & & & & & & &1& \\
        $\mathbf{10} $&$\ydiagram{2}$  &$\ydiagram{1,1}$&$3 $&2&1&1& & & & & & & &1& &1\\
        $\mathbf{14} $&$\ydiagram{2,2}$&$\ydiagram{2}$  &$5 $&2&1&1&1& & & & & &2& & & \\
        $\mathbf{30} $&$\ydiagram{3,3}$&$\ydiagram{3}$  &$9 $&2&2&1&1&1&1& & & & & &2& \\
        $\mathbf{35} $&$\ydiagram{3,1}$&$\ydiagram{2,1}$&$6 $&3&3&2&1&1& & & & & &1&1&1\\
        $\mathbf{35}'$&$\ydiagram{4}$  &$\ydiagram{2,2}$&$8 $&3&2&2&1&1& &1& & &2& &1& \\
        $\mathbf{55} $&$\ydiagram{4,4}$&$\ydiagram{4}$  &$14$&3&2&2&2&1&1&1&1&1&3& & & \\
        \hline\hline
    \end{tabular}
    \label{tbl:rep_deg}
\end{table*}

\begin{table*}[p]
    \centering
    \setlength{\tabcolsep}{5pt}
    \caption{The scaling dimensions of several operators at different $V/U$ and system size $N_\mathrm{orb}$ calibrated by the scaling dimension of the symmetry current $\Delta_J=2$. The quantum numbers $(\ell,\mathscr{P},\mathrm{rep.})$ are given in the bracket.
    (a) the conserved stress tensor $\mathscr{T}^{\mu\nu}$, fixed to be $3$ by conformal symmetry;
    (b) the difference of scaling dimension of $\phi$ and its descendant $\Delta_{\partial^\mu\phi}-\Delta_\phi$, fixed to be $1$ by conformal symmetry;
    (c) the difference of scaling dimension of $\phi$ and its descendant $\Delta_{\partial^\mu\partial^\nu\phi}-\Delta_\phi$, fixed to be $2$ by conformal symmetry;
    (d) the descendant $\epsilon^{\mu\nu\rho}\partial_\nu J_\rho$ of the conserved current $J^\mu$, fixed to be $3$ by conformal symmetry;
    (e) the descendant $\partial^\mu J^\nu$ of the conserved current $J^\mu$, fixed to be $3$ by conformal symmetry;
    (f) the difference of scaling dimension of $T$ and its descendant $\Delta_{\partial^\mu T}-\Delta_T$, fixed to be $1$ by conformal symmetry.}
    \begin{tabular}{r|lllll|lllll|lllll}
        \hline\hline
        \multirow{2}{*}{$V/U$}&\multicolumn{15}{c}{$N_\mathrm{orb}$}\\
        &10&9&8&7&6&10&9&8&7&6&10&9&8&7&6\\
        \hline
        &\multicolumn{5}{c}{\textbf{a}. $\mathscr{T}^{\mu\nu},(2,+,\mathbf{1})$}&\multicolumn{5}{|c|}{\textbf{b}. $\partial^\mu\phi,(2,-,\mathbf{5})-\phi$}&\multicolumn{5}{c}{\textbf{c}. $\partial^\mu\partial^\nu\phi,(1,-,\mathbf{5})-\phi$}\\
        \hline
        $0.3$ & $3.404$ & $3.413$ & $3.427$ & $3.452$ & $3.493$ & $1.403$ & $1.399$ & $1.395$ & $1.393$ & $1.392$ & $2.872$ & $2.856$ & $2.833$ & $2.803$ & $2.760$ \\
        $0.7$ & $3.084$ & $3.080$ & $3.079$ & $3.079$ & $3.086$ & $1.155$ & $1.152$ & $1.149$ & $1.146$ & $1.143$ & $2.196$ & $2.178$ & $2.155$ & $2.126$ & $2.086$ \\
        $0.9$ & $3.012$ & $3.005$ & $2.997$ & $2.989$ & $2.983$ & $1.104$ & $1.100$ & $1.096$ & $1.092$ & $1.086$ & $2.052$ & $2.031$ & $2.004$ & $1.970$ & $1.926$ \\
        $1.0$ & $2.986$ & $2.976$ & $2.965$ & $2.954$ & $2.943$ & $1.086$ & $1.082$ & $1.078$ & $1.072$ & $1.066$ & $2.000$ & $1.977$ & $1.948$ & $1.912$ & $1.865$ \\
        $1.5$ & $2.898$ & $2.880$ & $2.859$ & $2.834$ & $2.802$ & $1.033$ & $1.027$ & $1.020$ & $1.011$ & $0.999$ & $1.837$ & $1.806$ & $1.768$ & $1.723$ & $1.665$ \\
        $3.0$ & $2.788$ & $2.758$ & $2.720$ & $2.674$ & $2.615$ & $0.983$ & $0.974$ & $0.962$ & $0.947$ & $0.928$ & $1.661$ & $1.618$ & $1.568$ & $1.506$ & $1.431$ \\
        $10.0$& $2.678$ & $2.634$ & $2.579$ & $2.511$ & $2.427$ & $0.951$ & $0.938$ & $0.922$ & $0.901$ & $0.875$ & $1.525$ & $1.471$ & $1.408$ & $1.332$ & $1.239$ \\
        \hline
        &\multicolumn{5}{c}{\textbf{d}. $\epsilon^{\mu\nu\rho}\partial_\nu J_\rho,(1,-,\mathbf{10})$}&\multicolumn{5}{|c|}{\textbf{e}. $\partial^\mu J^\nu,(2,+,\mathbf{10})$}&\multicolumn{5}{c}{\textbf{f}. $\partial^\mu T,(1,+,\mathbf{14})-T$}\\
        \hline
        $0.3$ & $3.313$ & $3.294$ & $3.270$ & $3.241$ & $3.206$ & $3.473$ & $3.457$ & $3.432$ & $3.396$ & $3.343$ & $1.394$ & $1.389$ & $1.383$ & $1.378$ & $1.373$ \\
        $0.7$ & $3.099$ & $3.093$ & $3.084$ & $3.074$ & $3.060$ & $3.031$ & $3.016$ & $2.995$ & $2.967$ & $2.927$ & $1.145$ & $1.139$ & $1.132$ & $1.123$ & $1.111$ \\
        $0.9$ & $3.064$ & $3.059$ & $3.053$ & $3.045$ & $3.034$ & $2.945$ & $2.927$ & $2.905$ & $2.875$ & $2.835$ & $1.090$ & $1.083$ & $1.074$ & $1.062$ & $1.047$ \\
        $1.0$ & $3.052$ & $3.048$ & $3.042$ & $3.035$ & $3.026$ & $2.914$ & $2.896$ & $2.872$ & $2.841$ & $2.800$ & $1.070$ & $1.062$ & $1.052$ & $1.040$ & $1.023$ \\
        $1.5$ & $3.020$ & $3.016$ & $3.012$ & $3.006$ & $2.999$ & $2.821$ & $2.798$ & $2.770$ & $2.735$ & $2.690$ & $1.007$ & $0.996$ & $0.982$ & $0.965$ & $0.942$ \\
        $3.0$ & $2.988$ & $2.985$ & $2.980$ & $2.975$ & $2.968$ & $2.725$ & $2.696$ & $2.661$ & $2.619$ & $2.566$ & $0.938$ & $0.923$ & $0.903$ & $0.878$ & $0.846$ \\
        $10.0$& $2.962$ & $2.956$ & $2.950$ & $2.944$ & $2.936$ & $2.652$ & $2.618$ & $2.577$ & $2.527$ & $2.467$ & $0.886$ & $0.866$ & $0.840$ & $0.808$ & $0.766$ \\
        \hline\hline
    \end{tabular}
    \label{tbl:spec_scan}
\end{table*}

\begin{table*}[p]
    \setlength{\tabcolsep}{6pt}
    \centering\caption{The full low lying states with $\ell\leq 3$ and $\Delta<5.5$ measured at $N_\mathrm{orb}=10$ and $V/U=0.9437$, organized into different representations and conformal multiplets. The ``P'' and ``D'' in the last column denote the identified primaries and their descendants. The short dash divides multiplets and the line divides representations.}
    \label{tbl:full_spec}
    \begin{tabular}{ccccc|}
        \hline\hline
        $\ell$&$\mathscr{P}$&rep.&$\Delta$&\\
        \hline
        $0$& $+$  & $\mathbf{1}$   & $0.000$ &   \\ \cline{4-4}
        $0$& $+$  & $\mathbf{1}$   & $2.831$ & P \\
        $1$& $+$  & $\mathbf{1}$   & $3.911$ & D \\
        $2$& $+$  & $\mathbf{1}$   & $4.713$ & D \\
        $0$& $+$  & $\mathbf{1}$   & $5.055$ & D \\ \cline{4-4}
        $2$& $+$  & $\mathbf{1}$   & $3.000$ & P \\
        $3$& $+$  & $\mathbf{1}$   & $3.802$ & D \\
        $2$& $-$  & $\mathbf{1}$   & $4.014$ & D \\
        $3$& $-$  & $\mathbf{1}$   & $4.700$ & D \\
        $2$& $+$  & $\mathbf{1}$   & $4.945$ & D \\ \cline{4-4}
        $0$& $-$  & $\mathbf{1}$   & $5.338$ & P \\ \cline{4-4}
        $2$& $+$  & $\mathbf{1}$   & $4.919$ &   \\
        $3$& $+$  & $\mathbf{1}$   & $5.207$ &   \\ \hline
        $0$& $-$  & $\mathbf{5}$   & $0.585$ & P \\
        $1$& $-$  & $\mathbf{5}$   & $1.681$ & D \\
        $2$& $-$  & $\mathbf{5}$   & $2.614$ & D \\
        $0$& $-$  & $\mathbf{5}$   & $2.956$ & D \\
        $3$& $-$  & $\mathbf{5}$   & $3.308$ & D \\
        $1$& $-$  & $\mathbf{5}$   & $3.950$ & D \\
        $2$& $-$  & $\mathbf{5}$   & $4.732$ & D \\
        $0$& $-$  & $\mathbf{5}$   & $5.059$ & D \\ \cline{4-4}
        $2$& $-$  & $\mathbf{5}$   & $3.845$ & P \\
        $3$& $-$  & $\mathbf{5}$   & $4.664$ & D \\
        $2$& $+$  & $\mathbf{5}$   & $4.858$ & D \\
        $1$& $-$  & $\mathbf{5}$   & $5.082$ & D \\ \cline{4-4}
        $2$& $+$  & $\mathbf{5}$   & $3.986$ & P \\
        $3$& $+$  & $\mathbf{5}$   & $4.726$ & D \\
        $2$& $-$  & $\mathbf{5}$   & $4.903$ & D \\
        $1$& $+$  & $\mathbf{5}$   & $5.047$ & D \\ \cline{4-4}
        $0$& $-$  & $\mathbf{5}$   & $4.325$ & P \\
        $1$& $-$  & $\mathbf{5}$   & $5.330$ & D \\ \cline{4-4}
        $1$& $-$  & $\mathbf{5}$   & $4.501$ & P \\
        $2$& $-$  & $\mathbf{5}$   & $5.464$ & D \\
        \hline\hline
    \end{tabular}
    \!\!\!
    \begin{tabular}{ccccc|}
        \hline\hline
        $\ell$&$\mathscr{P}$&rep.&$\Delta$&\\
        \hline
        $3$& $-$  & $\mathbf{5}$   & $4.576$ &   \\
        $3$& $-$  & $\mathbf{5}$   & $5.249$ &   \\
        $2$& $-$  & $\mathbf{5}$   & $5.381$ &   \\
        $3$& $-$  & $\mathbf{5}$   & $5.382$ &   \\ \hline
        $1$& $+$  & $\mathbf{10}$  & $2.000$ & P \\
        $2$& $+$  & $\mathbf{10}$  & $2.931$ & D \\
        $1$& $-$  & $\mathbf{10}$  & $3.059$ & D \\
        $3$& $+$  & $\mathbf{10}$  & $3.628$ & D \\
        $2$& $-$  & $\mathbf{10}$  & $3.911$ & D \\
        $1$& $+$  & $\mathbf{10}$  & $4.081$ & D \\
        $3$& $-$  & $\mathbf{10}$  & $4.521$ & D \\
        $2$& $+$  & $\mathbf{10}$  & $4.857$ & D \\
        $1$& $-$  & $\mathbf{10}$  & $5.056$ & D \\ \cline{4-4}
        $1$& $+$  & $\mathbf{10}$  & $3.164$ & P \\
        $2$& $+$  & $\mathbf{10}$  & $4.064$ & D \\
        $1$& $-$  & $\mathbf{10}$  & $4.216$ & D \\
        $0$& $+$  & $\mathbf{10}$  & $4.304$ & D \\
        $3$& $+$  & $\mathbf{10}$  & $4.711$ & D \\
        $2$& $-$  & $\mathbf{10}$  & $5.051$ & D \\
        $1$& $+$  & $\mathbf{10}$  & $5.206$ & D \\ \cline{4-4}
        $3$& $+$  & $\mathbf{10}$  & $4.215$ & P \\
        $3$& $-$  & $\mathbf{10}$  & $5.103$ & D \\
        $2$& $+$  & $\mathbf{10}$  & $5.397$ & D \\ \cline{4-4}
        $3$& $-$  & $\mathbf{10}$  & $4.418$ &   \\
        $1$& $+$  & $\mathbf{10}$  & $4.515$ &   \\
        $2$& $-$  & $\mathbf{10}$  & $4.895$ &   \\
        $1$& $-$  & $\mathbf{10}$  & $5.144$ &   \\
        $3$& $+$  & $\mathbf{10}$  & $5.189$ &   \\
        $1$& $+$  & $\mathbf{10}$  & $5.239$ &   \\
        $3$& $+$  & $\mathbf{10}$  & $5.316$ &   \\
        $3$& $+$  & $\mathbf{10}$  & $5.408$ &   \\
        $3$& $+$  & $\mathbf{10}$  & $5.497$ &   \\ \hline
        $0$& $+$  & $\mathbf{14}$  & $1.458$ & P \\
        \hline\hline
    \end{tabular}
    \!\!\!
    \begin{tabular}{ccccc|}
        \hline\hline
        $\ell$&$\mathscr{P}$&rep.&$\Delta$&\\
        \hline
        $1$& $+$  & $\mathbf{14}$  & $2.538$ & D \\
        $2$& $+$  & $\mathbf{14}$  & $3.446$ & D \\
        $0$& $+$  & $\mathbf{14}$  & $3.753$ & D \\
        $3$& $+$  & $\mathbf{14}$  & $4.191$ & D \\
        $1$& $+$  & $\mathbf{14}$  & $4.659$ & D \\
        $2$& $+$  & $\mathbf{14}$  & $5.477$ & D \\ \cline{4-4}
        $2$& $+$  & $\mathbf{14}$  & $3.333$ & P \\
        $3$& $+$  & $\mathbf{14}$  & $4.038$ & D \\
        $2$& $-$  & $\mathbf{14}$  & $4.349$ & D \\
        $1$& $+$  & $\mathbf{14}$  & $4.720$ & D \\
        $3$& $-$  & $\mathbf{14}$  & $5.028$ & D \\
        $2$& $+$  & $\mathbf{14}$  & $5.232$ & D \\ \cline{4-4}
        $0$& $+$  & $\mathbf{14}$  & $4.351$ & P \\
        $1$& $+$  & $\mathbf{14}$  & $5.338$ & D \\ \cline{4-4}
        $2$& $+$  & $\mathbf{14}$  & $4.887$ &   \\
        $2$& $-$  & $\mathbf{14}$  & $5.299$ &   \\
        $2$& $+$  & $\mathbf{14}$  & $5.376$ &   \\
        $3$& $-$  & $\mathbf{14}$  & $5.500$ &   \\ \hline
        $1$& $-$  & $\mathbf{35}$  & $3.031$ & P \\
        $2$& $-$  & $\mathbf{35}$  & $3.953$ & D \\
        $1$& $+$  & $\mathbf{35}$  & $4.102$ & D \\
        $0$& $-$  & $\mathbf{35}$  & $4.244$ & D \\
        $2$& $+$  & $\mathbf{35}$  & $4.936$ & D \\
        $3$& $-$  & $\mathbf{35}$  & $5.015$ & D \\
        $1$& $-$  & $\mathbf{35}$  & $5.214$ & D \\ \cline{4-4}
        $2$& $-$  & $\mathbf{35}$  & $3.559$ & P \\
        $3$& $-$  & $\mathbf{35}$  & $4.334$ & D \\
        $1$& $-$  & $\mathbf{35}$  & $4.448$ & D \\
        $2$& $+$  & $\mathbf{35}$  & $4.541$ & D \\
        $3$& $+$  & $\mathbf{35}$  & $5.446$ & D \\
        $2$& $-$  & $\mathbf{35}$  & $5.454$ & D \\
        $1$& $+$  & $\mathbf{35}$  & $5.476$ & D \\ \cline{4-4}
        $3$& $-$  & $\mathbf{35}$  & $4.622$ & P \\
        \hline\hline
    \end{tabular}
    \!\!\!
    \begin{tabular}{ccccc}
        \hline\hline
        $\ell$&$\mathscr{P}$&rep.&$\Delta$&\\
        \hline
        $2$& $-$  & $\mathbf{35}$  & $5.305$ & D \\
        $3$& $+$  & $\mathbf{35}$  & $5.336$ & D \\
        $1$& $-$  & $\mathbf{35}$  & $4.734$ &   \\
        $2$& $-$  & $\mathbf{35}$  & $4.787$ &   \\
        $2$& $+$  & $\mathbf{35}$  & $4.805$ &   \\
        $1$& $+$  & $\mathbf{35}$  & $4.839$ &   \\
        $1$& $-$  & $\mathbf{35}$  & $5.121$ &   \\
        $3$& $+$  & $\mathbf{35}$  & $5.200$ &   \\ \hline
        $2$& $+$  & $\mathbf{35'}$ & $4.692$ & P \\
        $2$& $-$  & $\mathbf{35'}$ & $5.169$ & D \\
        $3$& $+$  & $\mathbf{35'}$ & $5.456$ & D \\ \cline{4-4}
        $0$& $+$  & $\mathbf{35'}$ & $4.885$ &   \\ \hline
        $0$& $-$  & $\mathbf{30}$  & $2.571$ & P \\
        $1$& $-$  & $\mathbf{30}$  & $3.621$ & D \\
        $2$& $-$  & $\mathbf{30}$  & $4.519$ & D \\
        $0$& $-$  & $\mathbf{30}$  & $4.803$ & D \\
        $3$& $-$  & $\mathbf{30}$  & $5.063$ & D \\ \cline{4-4}
        $2$& $-$  & $\mathbf{30}$  & $4.400$ & P \\
        $3$& $-$  & $\mathbf{30}$  & $5.284$ & D \\
        $2$& $+$  & $\mathbf{30}$  & $5.407$ & D \\ \cline{4-4}
        $3$& $-$  & $\mathbf{30}$  & $5.013$ &   \\ \hline
        $1$& $+$  & $\mathbf{81}$  & $4.295$ & P \\
        $2$& $+$  & $\mathbf{81}$  & $5.186$ & D \\
        $1$& $-$  & $\mathbf{81}$  & $5.340$ & D \\
        $0$& $+$  & $\mathbf{81}$  & $5.471$ & D \\ \cline{4-4}
        $2$& $+$  & $\mathbf{81}$  & $4.792$ &   \\
        $3$& $+$  & $\mathbf{81}$  & $5.185$ &   \\ \hline
        $0$& $+$  & $\mathbf{55}$  & $3.895$ & P \\
        $1$& $+$  & $\mathbf{55}$  & $4.901$ & D \\ \hline
        $0$& $-$  & $\mathbf{91}$  & $5.406$ & P \\
        &&&&\\
        &&&&\\
        &&&&\\
        \hline\hline
    \end{tabular}
\end{table*}

Specifically, for the Lorentz polarization, we consider eigenstates of $\hat{L}^2$ and $\hat{L}^z$ labeled by $l$ and $m$ and pick out $m=0$ components. For $\ell\leq 2$, the non-zero components of the polarizations are chosen as
\begin{align}
    n_{(l,m)=(0,0)}&=1\nonumber\\
    n_{(l,m)=(1,0)}^z&=1\nonumber\\
    n_{(l,m)=(2,0)}^{zz}&=2&n_{(l,m)=(2,0)}^{xx}&=n_{(l,m)=(2,0)}^{yy}=-1.
\end{align}

\clearpage

For the $\mathrm{SO}(5)$ polarization, we consider eigenstates of $\hat{\sigma}_1$ and  $\hat{\sigma}_2$. The basis of $\gamma$-matrices are taken as
\begin{equation}
    \gamma^{1,\dots,5}=\{\mathbb{I}\otimes\tau^x,\mathbb{I}\otimes\tau^z,\sigma^x\otimes\tau^y,\sigma^y\otimes\tau^y,\sigma^z\otimes\tau^y\},
\end{equation}
and the polarizations are determined by
\begin{equation}
    [e_{ij\dots}^{(\sigma_1,\sigma_2)}\gamma^i\otimes\gamma^j\otimes\dots,\hat{\sigma}^\alpha]=\sigma_\alpha e_{ij\dots}^{(\sigma_1,\sigma_2)}\gamma^i\otimes\gamma^j\otimes\dots,
\end{equation}
where $\alpha=1,2$. We pick out the $(\sigma_1,\sigma_2)=(0,0),(1,1)$ components of the vector representation and the $(\sigma_1,\sigma_2)=(1,1)$ components of the symmetric and antisymmetric rank-2 tensor representations. The non-zero components of the polarizations are chosen as
\begin{align}
    e^{\mathrm{S},(\sigma_1,\sigma_2)=(0,0)}&=1\nonumber\\
    e^{\mathrm{V},(\sigma_1,\sigma_2)=(0,0)}_2&=1\nonumber\\
    e^{\mathrm{V},(\sigma_1,\sigma_2)=(1,1)}_4&=1/\sqrt{2}&e^{\mathrm{V},(\sigma_1,\sigma_2)=(1,1)}_3&=-i/\sqrt{2}\nonumber\\
    e^{\mathrm{A},(\sigma_1,\sigma_2)=(1,1)}_{24}&=1/\sqrt{2}&e^{\mathrm{A},(\sigma_1,\sigma_2)=(1,1)}_{23}&=-i/\sqrt{2}\nonumber\\
    e^{\mathrm{T},(\sigma_1,\sigma_2)=(1,1)}_{24}&=1/\sqrt{2}&e^{\mathrm{T},(\sigma_1,\sigma_2)=(1,1)}_{23}&=-i/\sqrt{2}.
\end{align}

Hence, the normalizing factors are taken as
\begin{align}
    \alpha(S_{(l,m)=(0,0)}^{(\sigma_1,\sigma_2)=(0,0)})&=1&\alpha(\phi_{(l,m)=(0,0)}^{(\sigma_1,\sigma_2)=(0,0)})&=1\nonumber\\
    \alpha(\phi_{(l,m)=(0,0)}^{(\sigma_1,\sigma_2)=(1,1)})&=1&\alpha(T_{(l,m)=(0,0)}^{(\sigma_1,\sigma_2)=(1,1)})&=\sqrt{2}&\nonumber\\
    \alpha(J_{(l,m)=(1,0)}^{(\sigma_1,\sigma_2)=(1,1)})&=4&\alpha(\mathscr{T}_{(l,m)=(2,0)}^{(\sigma_1,\sigma_2)=(0,0)})&=\sqrt{8/27}.
\end{align}

With this, the 3-pt functions in general can be written as
\begin{widetext}
\begin{multline}
    \langle(\Phi_1)_\mathbf{n_1}^\mathbf{e_1}|(\Phi_2)_\mathbf{n_2}^\mathbf{e_2}(\theta,\varphi)|(\Phi_3)_\mathbf{n_3}^\mathbf{e_3}\rangle=\alpha^*((\Phi_1)_\mathbf{n_1}^\mathbf{e_1})\alpha((\Phi_3)_\mathbf{n_3}^\mathbf{e_3})\lim_{x\to\infty}x^{2\Delta_1}(n_1^*)^{\mu''_1\dots\mu''_\ell}n_2^{\mu'_1\dots\mu'_\ell}n_3^{\mu_1\dots\mu_\ell}e^*_{1,i''j''\dots}e_{2,i'j'\dots}e_{3,ij\dots}\\
    \times I_{\mu''_1}{}^{\nu_1}(x)\dots I_{\mu''_\ell}{}^{\nu_\ell}(x)\left\langle(\Phi_1)_{\nu_1\dots\nu_\ell}^{i''j''\dots}(x)(\Phi_2)_{\mu'_1\dots\mu'_\ell}^{i'j'\dots}(\theta,\varphi)(\Phi_3)_{\mu_1\dots\mu_\ell}^{ij\dots}(0)\right\rangle.
\end{multline}
Specifically
\begin{align}
    \langle\phi_{(l,m)=(0,0)}^{(\sigma_1,\sigma_2)=(0,0)}|\phi^{(\sigma_1,\sigma_2)=(0,0)}(\theta,\varphi)|S_{(l,m)=(0,0)}^{(\sigma_1,\sigma_2)=(0,0)}\rangle&=R^{-\Delta_\phi}f_{\phi\phi S}\nonumber\\
    \langle\phi_{(l,m)=(0,0)}^{(\sigma_1,\sigma_2)=(1,1)}|\phi^{(\sigma_1,\sigma_2)=(0,0)}(\theta,\varphi)|J_{(l,m)=(1,0)}^{(\sigma_1,\sigma_2)=(1,1)}\rangle&=R^{-\Delta_\phi}f_{\phi\phi J}\cos\theta\nonumber\\
    \langle\phi_{(l,m)=(0,0)}^{(\sigma_1,\sigma_2)=(1,1)}|\phi^{(\sigma_1,\sigma_2)=(0,0)}(\theta,\varphi)|T_{(l,m)=(0,0)}^{(\sigma_1,\sigma_2)=(1,1)}\rangle&=\frac{1}{\sqrt{2}}R^{-\Delta_\phi}f_{\phi\phi T}\nonumber\\
    \langle\phi_{(l,m)=(0,0)}^{(\sigma_1,\sigma_2)=(0,0)}|\phi^{(\sigma_1,\sigma_2)=(0,0)}(\theta,\varphi)|\mathscr{T}_{(l,m)=(2,0)}^{(\sigma_1,\sigma_2)=(0,0)}\rangle&=\frac{1}{\sqrt{6}}R^{-\Delta_\phi}f_{\phi\phi\mathscr{T}}(1+3\cos 2\theta).
\end{align}
\end{widetext}

We then integrate out the angular dependence by taking the angular momentum component

\begin{equation}
    \hat{\phi}_{lm}=\int\frac{\sin\theta\,\mathrm{d}\theta\,\mathrm{d}\varphi}{4\pi}\bar{Y}_{lm}(\theta,\varphi)\hat{\phi}(\theta,\varphi).
\end{equation}
In our calculation, we use the density operator $\hat{n}$ instead of $\hat{\phi}$. To the leading order
\begin{equation}
    \hat{n}(\theta,\varphi)=\alpha_\phi\hat{\phi}+\dots
\end{equation}
and the coefficient can be accessed by
\begin{equation}
    \langle\phi|\hat{n}(\theta,\varphi)|0\rangle=\alpha_\phi R^{-\Delta_\phi}(1+\mathscr{O}(R^{-2})).
\end{equation}
The subleading contribution comes from the descendants of $\phi$ and other multiplets in the same sector. 

Hence, to the leading order
\begin{align}
    f_{\phi\phi S}&=\frac{\langle\phi_{(l,m)=(0,0)}^{(\sigma_1,\sigma_2)=(0,0)}|n_{(l,m)=(0,0)}^{(\sigma_1,\sigma_2)=(0,0)}|S_{(l,m)=(0,0)}^{(\sigma_1,\sigma_2)=(0,0)}\rangle}{\langle\phi_{(l,m)=(0,0)}^{(\sigma_1,\sigma_2)=(0,0)}|n^{(\sigma_1,\sigma_2)=(0,0)}_{(l,m)=(0,0)}|0\rangle}\nonumber\\
    f_{\phi\phi J}&=\sqrt{3}\frac{\langle\phi_{(l,m)=(0,0)}^{(\sigma_1,\sigma_2)=(1,1)}|n_{(l,m)=(1,0)}^{(\sigma_1,\sigma_2)=(0,0)}|J_{(l,m)=(1,0)}^{(\sigma_1,\sigma_2)=(1,1)}\rangle}{\langle\phi_{(l,m)=(0,0)}^{(\sigma_1,\sigma_2)=(0,0)}|n^{(\sigma_1,\sigma_2)=(0,0)}_{(l,m)=(0,0)}|0\rangle}\nonumber\\
    f_{\phi\phi T}&=\sqrt{2}\frac{\langle\phi_{(l,m)=(0,0)}^{(\sigma_1,\sigma_2)=(1,1)}|n_{(l,m)=(0,0)}^{(\sigma_1,\sigma_2)=(0,0)}|T_{(l,m)=(0,0)}^{(\sigma_1,\sigma_2)=(1,1)}\rangle}{\langle\phi_{(l,m)=(0,0)}^{(\sigma_1,\sigma_2)=(0,0)}|n^{(\sigma_1,\sigma_2)=(0,0)}_{(l,m)=(0,0)}|0\rangle}\nonumber\\
    f_{\phi\phi\mathscr{T}}&=\sqrt{\frac{15}{8}}\frac{\langle\phi_{(l,m)=(0,0)}^{(\sigma_1,\sigma_2)=(0,0)}|n_{(l,m)=(2,0)}^{(\sigma_1,\sigma_2)=(0,0)}|\mathscr{T}_{(l,m)=(2,0)}^{(\sigma_1,\sigma_2)=(0,0)}\rangle}{\langle\phi_{(l,m)=(0,0)}^{(\sigma_1,\sigma_2)=(0,0)}|n^{(\sigma_1,\sigma_2)=(0,0)}_{(l,m)=(0,0)}|0\rangle}.
\end{align}

\clearpage

\end{document}